 \documentclass[twocolumn,trackchanges,usenames,dvipsnames]{aastex631}
\usepackage[english]{babel}

\usepackage{amsmath,amssymb}
\usepackage{graphicx}
\usepackage{xcolor}
\usepackage{latexsym}
\usepackage{dcolumn}
\usepackage{amsmath}
\usepackage{epsf}
\usepackage{float}
\usepackage{enumerate}
\usepackage{enumitem}
\usepackage{graphicx}
\usepackage{dcolumn}
\usepackage{bm}
\definecolor{mynavy}{RGB}{18,18,155}
\usepackage{soul}  
\setstcolor{red}
\usepackage{wasysym}
\usepackage{rotating}
\usepackage{multirow}
\usepackage{chemfig}
\usepackage{enumitem}
\usepackage{ulem,xcolor}

\usepackage{natbib}
\usepackage{chngcntr}

\begin{document}

\title{Observational prospects of double neutrons star mergers and their multi-messenger afterglows: LIGO discovery power, event rates and diversity}

\author{Maryam Aghaei Abchouyeh}
\affiliation{Department of Physics and Astronomy, Sejong University, 98 Gunja-Dong, Gwangjin-gu, Seoul 143-747, South Korea}

\author{Maurice H.P.M. van Putten}
\altaffiliation{Corresponding author, mvp@sejong.ac.kr}
\affiliation{Department of Physics and Astronomy, Sejong University, 98 Gunja-Dong, Gwangjin-gu, Seoul 143-747, South Korea}

\author{Lorenzo Amati}
\affiliation{INAF-IASF Bologna, via P. Gobetti, 101, I-40129 Bologna, Italy}





\begin{abstract}
The double neutron star (DNS) merger event GW170817 signifies the first multimessenger (MM) event with electromagnetic-gravitational (EM-GW) observations. LIGO-Virgo-KAGRA observational runs O4-5 promise to detect similar events and as yet unknown GW signals, which require confirmation in two or more detectors with comparable performance. To this end, we quantify duty cycles of comparable science quality of data in coincident H1L1-observations, further to seek consistent event rates of astrophysical transients in upcoming EM-GW surveys. Quite generally, discovery power scales with exposure time, sensitivity, and critically depends on the percentage of time when detectors operate at high quality. We quantify coincident duty cycles over a time-frequency domain $W\times B$, defined by segments of duration $W=8$s, motivated by a long-duration descending GW-chirp during GRB170817A, and the minimum detector noise over about $B=100-250\,$Hz. This detector yield factor satisfies $1\%-25 \%$ in S5-6 and O1-O3ab, significantly different from duty cycles of H1 and L1 individually with commensurable impact on consistency in event rates in EM-GW surveys. Significant gain in discovery power for signals whose frequency varies slowly in time may be derived from improving detector yield factors by deploying time-symmetric data analysis methods. For O4-5, these can yield improvements by factors up to $\mathcal{O}(10^5)$ relative to existing data and methods. Furthermore, the diversity of MM afterglows to DNS mergers may be greatest for systems similar to GW170817 but possibly less so for systems of substantially different mass such as GW190425. We summarize our findings with an outlook on EM-GW surveys during O4-5 and perspectives for next-generation GRB missions like {\sc THESEUS}.
\end{abstract}




\section{Introduction \label{Sec:In}}
Astronomers have since long waited for independent observation and identi!cation of astrophysical transients using multimessengers (MMs). These messengers include high- energy transients in the universe emitting, notably, cosmic rays, high-energy neutrinos, electromagnetic waves (EMs) from radio to -ray, and gravitational waves (GWs) \citep{Bart2017,Mesz2019}). GWs have negligible interaction with matter in the universe and therefore reach us essentially uninhibited making them stand out in this observa- tional approach.
In MM-astronomy (MMA), real-time alerts from different transient observatories are key to discovery \citep{Dorn2021} and there have been numerous efforts to achieve this goal during the past decades. The concept of MMA, although anticipated for centuries, is a relatively recent development. The first reported MM event is the observation of neutrinos associated with Supernova 1987A a few hours before the observation of visible light of the supernova shock breakout \citep{Arne1989}.
 
The next breakthrough was the gravitational-wave event GW170817 detected by LIGO-Virgo \cite{Abbott2017e} with an MM afterglow in a short -ray burst (SGRB) GRB170817A, identified by the Fermi-Gamma-Ray Burst Monitor (GBM) and INTErnational Gamma-Ray Astrophysics Laboratory (INTEGRAL; \citep{2017GCN.21506....1C,2017ApJ...848L..14G,2017Sci...358.1559K,2017ApJ...848L..15S,2018ApJ...852L..30P}). Importantly, GW170817 is accompanied by a kilonova AT2017gfo, the latter evidencing cosmic production of elements heavier than iron in the r-process \citep{Abbott2017d,Troj2017,Hall2017,Abbott2017g,Alex2017,Marg2017,Kasl2022}. While this event is expected to be also luminous in megaelectronvolt-neutrinos, such is not detectable from its extragalactic distance of about 40 Mpc. The nondetection of high-energy neutrinos may be due to seeing GRB170817A off- axis \citep{Albe2017}. GW170817 was the first gravitational- wave event with an EM counterpart, marking a second birth of MM-astronomy. In the same year, Ice-Cube detected an extremely high-energy neutrino event with -ray counterpart reported by \textit{Fermi}-LAT and the Major Atmospheric Gamma Imaging Cherenkov (MAGIC) telescope system from a blazar at the same location as the neutrino source \citep{CollabI2018}.

Furthering these developments comes from a network of observatories to explore cosmic transients at high precision including LIGO-Virgo, soon to be including KAGRA \citep{2021PTEP.2021eA103A}, ENGRAVE, GRANDMA, GROWTH, and GOTO \citep{Ackl2020,Anti2020a,Coug2019a,Gomp2020} \footnote{http://www.engrave-eso.org, https://grandma.ijclab.in2p3.fr}.  These observatories were collaborating during the third observational run (O3) of LIGO-Virgo aiming to detect the counterparts of GW triggered events \citep{Anti2020,Leva2020}. Nevertheless, no EM afterglows were detected during this run because candidate events were relatively distant with rather wide localization area \citep{Bart2017,Leva2020,Anti2020}. Needless to say, at present, the most exciting MM-events include the double neutron star (DNS) mergers, neutron star-black hole (NS-BH) mergers, and potentially core-collapse supernova (CC-SNe). While a few observational candidates exist for the first two, the last one remains enigmatic for their unknown central engine: a rotating neutron star or black hole \citep[e.g.][]{2010Natur.463..513S} with vastly distinct prospects on their potential GW emission in both energy and frequency \citep[cf.][]{Putten2019,Putten2022}. These key focal points motivate the deployment of a coherent and reliable network of observatories in the development of EM-GW surveys.

{Quite generally, also to set notation, {MMA transient events present radiation energies $\mathcal{E}_i$ across multiple energy channels} in transition to a remnant with mass-energy $E_R\le E_P$ produced by a progenitor with total mass-energy $E_P$. The $\mathcal{E}_i$ 
{satisfies} total energy conservation, i.e.:
\begin{equation}
    \label{EC} 
    \mathcal{E}_i\leq \sum_{k} \mathcal{E}_k \leq E_{P}-E_{R}\leq E_{P},
\end{equation}
where $k$ runs over the different radiation channels in GW, EM, neutrinos, mass-ejecta and magnetic winds. Among these, GW may be dominant while neutrino-emission and EM may take second and, respectively, third place \citep{Putten2003}.}

{Already, (\ref{EC}) is illustrated by the very first two MM transients SN1987A and GW170817 (Fig.\ref{fig:11}), both with leading energy emission outside the EM channel in neutrinos \citep{Hira1987,Kouv1993} and, respectively, GW \citep{Abbott2017}. Specializing to 
{compact binary coalecences (CBCs) of DNS and BH-NS}, \eqref{EC} translates to $\mathcal{E}_i\leq E_{CBC}-E_{R}\leq E_{CBC}$ with radiation output distributed over a {GRB, kilonova, GWs and neutrinos (unseen for GW170817)}. In particular, 
\begin{equation}
    \label{EC1} 
    \mathcal{E}_{GW}\leq E_{P}
\end{equation}
is a secure prior on potential 
{GW emission} following final coalescence. We expect this to be elucidated further in joint EM-GW surveys over the planned O4-5 observational runs.}

An important consideration in MM-observations is simultaneous observing time in a network {of detectors operating at comparable sensitivity and stability, enabling meaningful cross-correlations.} The concept of simultaneous observing time gains more significance considering the fact that no observatory is {necessarily all-sky and operating continuously at nominal sensitivity at any given moment}. 
{Among the various energy channels, GW-observations stand out to which the Universe is transparent, particularly distinct from EM-observations. 
Presently, the LVK-network of GW-detectors consists of the LIGO detectors H1 and L1 in the US, the Virgo detector in Italy, the KAGRA detector in Japan \citep{2021PTEP.2021eA103A} {and GEO600 in Germany \citep{Abbott2008b}. Shortly after the end of the last LVK observational run, KAGRA had a joint observational run with GEO600 \citep{Abbott2022}.} } 
The duty cycle of these LVK detectors and efficiency of searches close to nominal detector sensitivities are significant factors in the realization of their full discovery power - the probability of detecting a signal from an astrophysical transient in the Local Universe - especially so in a MM context of {new GW-signals hitherto unseen}. 

{With a focus on potential observations of a new GW-signal during O4-5, we here revisit and quantify the performance {of the two LIGO detectors for data-quality} in joint detector operations as part of an effort to connect a LIGO observations to existing EM-surveys. A first-time observation of a new signal, possibly at the threshold of detection of each individual detector, requires confirmation by two (or more) detectors. Quite generally, such requires two detectors operating at comparable sensitivities. For this reason, the present study considers H1 and L1, leaving Virgo to be included a future study.}

We begin this study with a revisit of duty cycles of individual and joint detector operations derived from the most sensitive frequency range for LIGO across the bandwidth
\begin{eqnarray}
B = 100-250\,\mbox{Hz}
\label{B}
\end{eqnarray}
for each run S5-6 and O1-O3ab.
{{$B$ is relevant in defining the minimum in detector noise amplitude}, while being largely free of lines. For an equal mass 
{DNS} merger, it captures about one-fourth of the total energy output up to the characteristic frequency of about 650 Hz before final coalescence \citep{Abbott2017f}. Hence, this frequency range is representative for the horizon distance of DNS mergers. 

By universality of astrophysical black holes across different mass scales, it is expected to be relevant to the prospective central engines of GRBs and energetic CC-SNe of type Ib/c alike, the latter broadly representing the progenitors of long GRBs \citep[cf.][]{2007PASP..119.1211F,2010NewAR..54..206L}. 
Scaling of the observed descending GW-chirp in frequency $f_{GW}$ by mass $M$ of the Kerr black hole powering GRB170817A \citep{Putten2022} gives
\begin{eqnarray}
200\, \mbox{Hz}\left(\frac{2.5M_\odot}{M}\right)\lesssim f_{GW}\lesssim 700\, \mbox{Hz}\left(\frac{2.5M_\odot}{M}\right).
\label{EQN_fGW1}
\end{eqnarray}
The scaling relation (\ref{EQN_fGW1}) defines an outlook for similarly powered CC-SNe, upon estimating the mass $M_{He}$ of the helium core prior to core-collapse and the mass-fraction $\xi$ thereof retained in collapse to a rotating black hole. From a He-core mass-function $dN/dM_{He}\propto e^{-M/M_s}$ over $M_c< M_{He} < 50M_\odot$ with cut-off $M_c\simeq 6M_\odot$ \citep{2015MNRAS.446.1213K}, we infer $\bar{M}_{He}\simeq 2 M_c$.
Thus, $\bar{M} = 2\xi M_c$ with 
$\xi\simeq 0.4$ for CC-SNe in close stellar binaries {produces} relatively rapidly rotating black holes \citep{2004ApJ...611L..81V}. As a result, we expect descending GW-chirps also from a fraction of CC-SNe extending over 
\begin{eqnarray}
100\,\mbox{Hz}\lesssim f_{GW} \lesssim 360\,\mbox{Hz}.
\label{EQN_fGW2}
\end{eqnarray}
{This frequency range (\ref{EQN_fGW2})}, similar to (\ref{B}), appears opportune given the above-mentioned minimum of LIGO detector strain noise amplitude.

{Accurate detector yield factors are essential in seeking consistency in event rates in EM-GW surveys. Results are particularly relevant for afterglows to DNS mergers. 
Considerable diversity in their MM afterglow emission is expected, both intrinsic and due to observational selection effects. Intrinsic diversity is expected as a function of total mass of a DNS system, a primary parameter defining the lifetime of the initial hyper-massive neutron star (HMNS) \citep{Lucca2020}. A GRB association such as GRB170817A to GW170817 is inevitably subject to small observed-to-true event ratios by their beaming factors. This problem continues to receive broad attention for both short and long GRBs
\citep[e.g.][]{Sala2015,Mand2018,Farah2020,Dich2020,Frye2022}.} 
We include these results in our outlook on O4-5. 

{In light of the above, we include an overview of LIGO observations S5-6 and O1-O3ab and their sensitivity, detector duty cycles and horizon distances (Appendix A). A primary motivation of the present work is probing for the central engines of extreme transient events such as cosmological GRBs associated with mergers, normal long GRBs and their parent population of energetic CC-SNe. Specifically, the present work focuses on searches for
\renewcommand{\labelenumi}{\alph{enumi})}
\begin{enumerate}
\item Unknown GW-signals from un-modeled sources requiring confirmation across two (or more) detectors with comparable performance in sensitivity;
\item Consistent event rates in EM-GW surveys on astrophysical transient events in the Local Universe.
\end{enumerate}

\begin{figure*}
\centering
\includegraphics[width=0.8\textwidth,trim=4 4 4 4,clip]{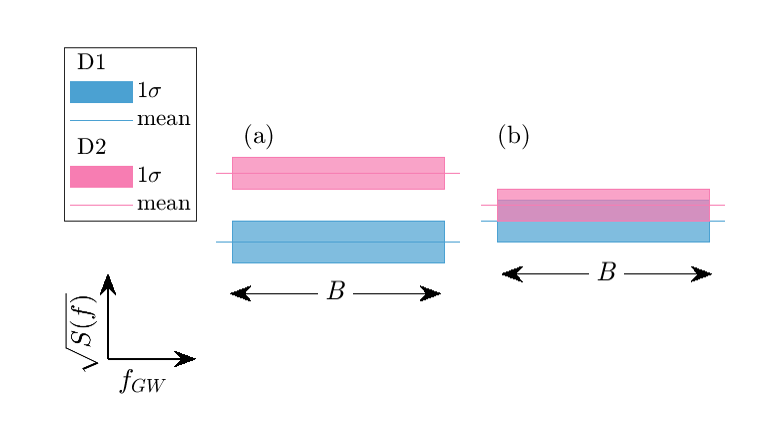}
\caption{\small {Detector duty cycles producing science quality data can be inferred from detector stain-noise spectra, $\sqrt{S(f)}$ versus $f_{GW}$ (left bottom). Science quality data is defined by stable and nominal spectra over segments of predefined duration $W$, evaluated by Welch method over subintervals of $w$, covering a choice of bandwidth $B$ \eqref{B}. Here we schematically show the definition of coincident duty cycle ($\mathcal{U}$) commonly used by LVK (a) and detector yield factor $y$ used in the present work (b). Solid lines and filled rectangles are representative for the mean of $\sqrt{S(f)}$ for a pair of detectors D1 and D2, and, respectively, the associated standard deviation ($\sigma$) therein. It generally depends on the choice of time-frequency domain $W\times B$. (a) requires stable performance for D1 and D2 individually with no regards to comparable performance. In (b), we require comparable performance highlighted by overlap shown within the bound \eqref{ms}. Evidently, inequality \eqref{yield} holds.}
\label{overlap}}
\end{figure*}

In preparing for O4-5, we approach (a-b) in various steps, starting with detector performance relevant to transient events with the potential to 
{produce} unknown GW-signals lasting several seconds as follows:
\begin{itemize}
\item Measuring comparable detector 
{performances} by statistics of high-quality data in coincident 
H1L1-operation over a time-frequency window $W\times B$ and the {duty cycle thereof (\S\ref{Sec:ADC1})}, where $W$ refers to a choice of duration of data-segments representative for long-duration GW-transients;
\item Enhancing discovery power when probing long-duration transient events over several seconds (\S\ref{Sec:IF}) using time-symmetric methods of analysis at sensitivities on par with CBC, satisfying (\ref{EC1}).
\item Diversity in MM-afterglows to DNS mergers as a function of total mass (\S\ref{Sec:DNS}), supported by progenitor mass-components from existing radio surveys,  illustrated by GW170817 and, possibly, GW190425. Additionally, we discuss the GRB211227A-kilonova assocation and DNS mass components.
\item Outlook on O4-5 focused on diversity in MM-afterglow emissions to mergers (\S\ref{Sec:Co}).
Discussed is further the planned GRB-mission THESEUS \citep{Amati2018,Amati2021}.
\end{itemize}
}

\section{LIGO Detector Performance} \label{Sec:ADC1} 

We revisit LIGO runs to estimate duty cycles of high-quality data in coincident detector operation at nominal behavior, {essential to searches of unknown GW-signals in probing} central engines of GRBs and CC-SNe in the Local Universe (e.g. \cite{Cowa2002,Putten2019}). 

We target 
{GW} transients with duration similar to the signals from {the central engine of GRB170817A \citep{Putten2022},
i.e., long duration transients extending over several seconds or more. Therefore, the primary focus of our analysis is on the segments of
\begin{equation}
    \label{sub-frame}
    W=8\ \mbox{seconds}.
\end{equation}
In particular, our focus is entirely separate from searches for BBH megers, whose durations are about one order of magnitude shorter.} 

While high-quality data from one detector {may suffice to probe mergers by their by now familiar ascending GW-chirps}, this is not the case for new sources. 
{To assess sensitivity for the latter, we perform a broad study over S5, S6, O1, O2, O3ab using above-mentioned frequency range (\ref{B}) as a proxy. According to Appendix A (Table \ref{T1}), results of this revisit are expected to be illustrative for upcoming observational runs O4-5.}

To this end, we extract statistically representative samples of {data-frames of} 4096 seconds (down-sampled to 4096\,Hz) by randomly selecting 112 central GPS times in each run. {Thus, a total of 672 randomly selected frames 
{are used} for analysis by the following steps:}

{{\it Step (a):}} Table \ref{T2}, column 2 reveals the existence of gaps \textit{gaps} in each run. 
  Listed are the number of available data following the randomly selected GPS times.

{{\it Step (b):}} {Available data} are down-selected to {paired H1L1-data. 
  In this process, we note there are some incomplete frames, not covering the whole duration of 4096 \,s and there are instances when only one of the detectors is in observing mode}. Notably, there is a probability of having {too many gaps, inhibiting extraction of spectra (and spectrograms).} These results are presented in column 3 of Table \ref{T2}, roughly equivalent to column 4 of Table \ref{T1}, i.e., the coincidence duty cycle of H1 and L1 ($\mathcal{U}$) during each run.

{{\it Step (c):}} 
  {Since a full data does not guarantee quality, we next check for the joint H1L1-data segments with at least one of the detectors collecting data without significant gaps (Column 4 of Table \ref{T2}).} 
 {{As our target signals are GWs transients with the duration in \eqref{sub-frame}, to do this analysis we split data into segments of 8 seconds for checking the quality of data.}} {We determine data-quality 
 {for each segment} according to detector sensitivity over $B$ by the Welch method applied to Fourier windows of size}
  \begin{equation}
  \label{window}
  w=W/2=4\ \mbox{s}. 
  \end{equation}

For S6 with $w=64$s 
{over the whole of 4096 s data}, our statistical results are in {quantitative agreement} with \cite{Putten2017}, in which $all$ 12728 frames were analyzed. This evidences that our random selection of 112 GPS times is sufficient. 

{{\it Step (d):}} {As a prerequisite for coincidence high-quality segments of H1L1-data,  both H1 and L1 must be free of significant gaps. The number of segments {that satisfy this condition is quite} small. The least and largest are found in S6 and O3a (S5), respectively (Table \ref{T2}, column 5). Notably absent is a discernible trend that might point to significant improvements of detectors in later runs.}
		\begin{table*}[]
			\footnotesize
			\begin{tabular}{|c||c|c|c|c|c|}
				\hline
				\hspace{0.1cm}	& {\bf Joint} & {\bf joint H1-L1 4096 s:}    & {\bf Segment}       & {\bf Joint}     &  {\bf Joint}   \\ 
				 	            &  {\bf H1-L1 data}   & {\bf Not all zero}     &  {\bf Densities}           & {\bf H1-L1 segments}           & {\bf H1-L1 segments at} \\ 
				\hspace{0.1cm}  &       	&                &              &           & {\bf high quality ($y$)} \\ \hline \hline
				 
				&    &  &  &  &     \\
				S5	& \pmb{$80\%$}  & \pmb{$78\%$} & \pmb{$77\%$} &  \pmb{$63\%$} &  \pmb{$13\%$}  \\ \hline
				
				&    &  &  &  &     \\
				S6	& \pmb{$40\%$ } & \pmb{$37\%$} & \pmb{$34\%$} &  \pmb{$20\%$} &  \pmb{$8\%$}  \\ \hline
					&    &  &  &  &       \\
				O1	& \pmb{$45\%$}  & \pmb{$42\%$} & \pmb{$41\%$} & \pmb{$34\%$} &  \pmb{$1\%$} \\ \hline
					&    &  &  &  &      \\
				O2	&\pmb{$49\%$}  & \pmb{$44\%$}  & \pmb{$42\%$} & \pmb{$36\%$} &  \pmb{$19\%$} \\ \hline
					&    &  &  &  &    \\
				O3a	& \pmb{$75\%$}  & \pmb{$70\%$}  & \pmb{$69\%$} & \pmb{$54\%$} &   \pmb{$14\%$}  \\ \hline
					&    &  &  &  &     \\
				O3b	& \pmb{$58\%$} & \pmb{$57\%$} & \pmb{$55\%$} & \pmb{$44\%$}  &  \pmb{$25\%$}  \\ \hline
				
			\end{tabular}
			\caption{\small {\textit{Summary of data-quality in LIGO runs }}. Results derived from statistics of 112 frames randomly selected by GPS times, evaluated for availability and quality of data. A segment of (8\,s) is considered in searches of long duration transients. Joint H1L1-operation at comparable sensitivities is crucial in discovery of unknown 
   {GW signals} from un-modeled sources, requiring coincident high-quality data. Column 2 lists gaps in the data. Deviation from $100\%$ indicates the fraction of GPS times in a run with no available data. Even when frames exist, they may not cover the whole 4096\,s. Column 3 indicates the fraction of joint full 
   {frame out of 112 (independent from target of search i.e., $w$-independent)}, roughly equivalent to Column 4 of Table \ref{T1}. 
   A full frame does not guarantee quality data or the absence of significant gaps. Gap statistics is $w$-dependent, i.e., it generally depends on search target. To this end, we 
   consider segments of duration $\Delta t=W=2w=8$ seconds for further analysis (\ref{sub-frame}). 
   Column 4 lists the number of joint H1L1-data segments in which at least one of the detectors is free of significant gaps. 
   The fraction of segments with no gaps listed in Column 5 significantly decreases compared to Column 4 with no evident trend with runs. 
   {Finally, column 6 lists joint H1L1-data segments 
   {both} in observing mode at high-quality, defined by $\sigma/\mu < 1$ simultaneously for both detectors across the frequency range (\ref{B}).} 
   {This measure can be further restricted or relaxed depending on search scope.} The results illustrate that the fraction of observational time with high-quality data simultaneously for both detectors of interest to un-modeled searches over $W=8\,$s is roughly $10-20\%$ with no significant improvement across the different runs.}
	\label{T2}
	\end{table*}

{{\it Step (e):}} 
{Finally, we consider both detectors in observing mode collecting 
high-quality data, defined when 
H1 and L1 show comparable nominal performance permitting cross-correlation of candidate observational events. Specifically, we consider
an overlap of 
{detectors} strain-noise amplitude within $0.5\sigma$, $1\sigma$ or $2\sigma$ band in above-mentioned most sensitive frequency interval and, correspondingly,}
\begin{equation}
    \label{ms}
    \dfrac{\sigma}{\mu} < \dfrac{1}{2}, 1\mbox{~or}\ 2,
\end{equation}
where $\mu$ is the mean value of the spectrum {(one-sided PSD)} based on Welch method and $\sigma$ is the standard deviation of the spectrum at each frequency  calculated over {two half-segments of FFT window size $w=4$s conform \eqref{window}}. {Fig. \ref{overlap} schematically shows the difference between the definition of high-quality data required in search of unknown GW signals, and coincidence duty cycle (used by LVK) which is defined based on individual stable performance in L1 and H1. }

\begin{figure*}[!htb]                        
\minipage{0.50\textwidth}
\includegraphics[width=\linewidth]{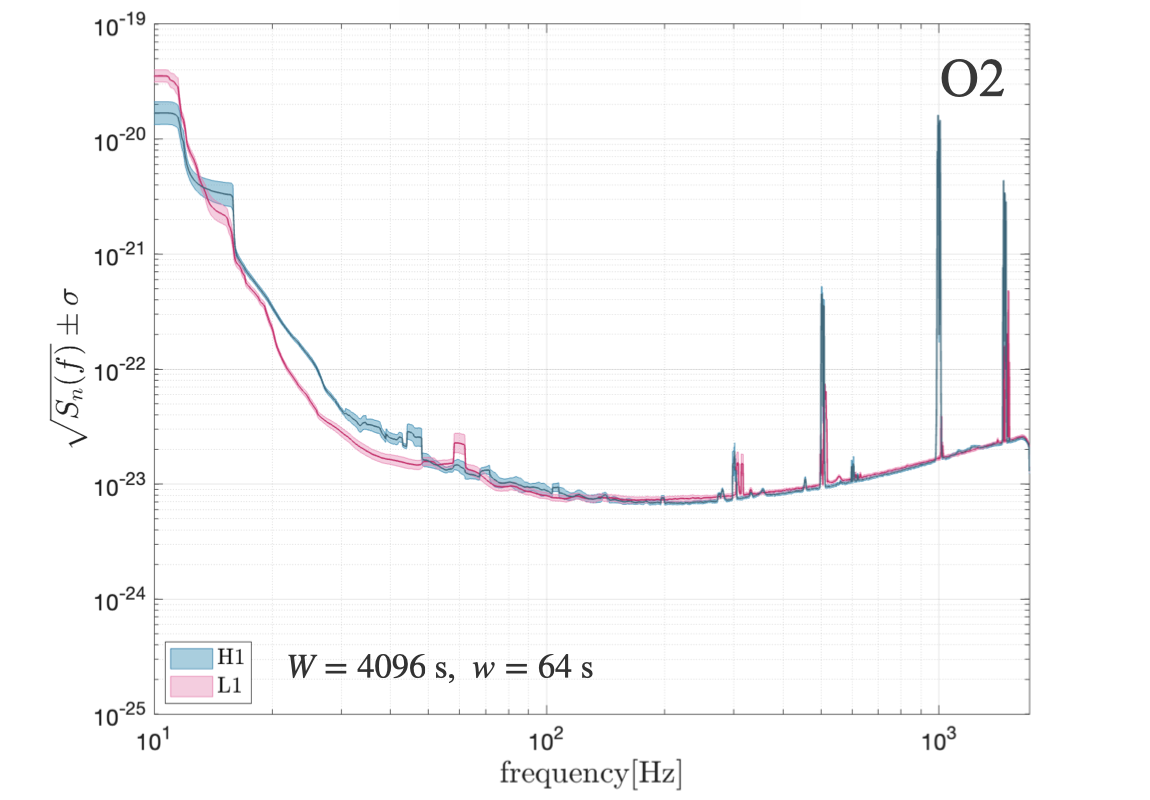}
\endminipage\hfill
\minipage{0.5\textwidth}
\includegraphics[width=\linewidth]{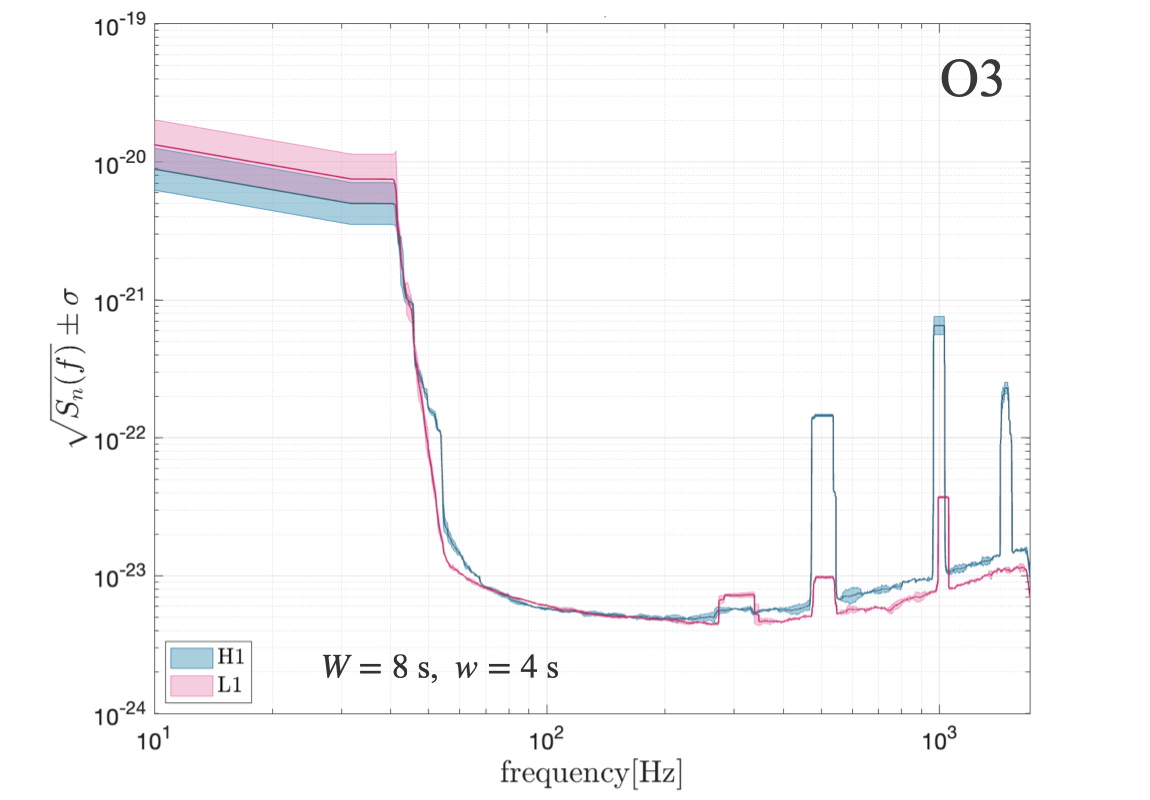}	
\endminipage\hfill
\caption{\small {{Sample spectra of O2 (left) and O3 (right) for the segments of $W=4096$\,s and $W=8$\,s, respectively.  
{These samples represent} high-quality data with small standard deviation for both H1 and L1. {Their nominal behavior shows} overlap within the most sensitive frequency interval of LIGO over about $B$ \eqref{B}. 
The right panel corresponds to our target in searches of un-modeled signals.  At lower frequencies seismic noise is dominant, while at higher frequencies, optical shot-noise is dominant. The difference in the line widths is due to different sizes of data segments and {smoothing by a moving average over frequency.}} }
\label{fig:12}}
\end{figure*} 

Fig. \ref{fig:12} shows an example of coincidence high-quality data-segment \eqref{sub-frame} in joint H1L1-operation with $\sigma/\mu<1$ over the most sensitive frequency interval (\ref{B}). Applying these restrictions on the remaining pair of segments, LIGO is found to be observing in high-quality regime generally up to $25\%$ (Table \ref{T2}, column 6). Hereafter, we refer to these numbers as the \textit{detector yield factor} ($y$) of each run, satisfying (Fig. \ref{fig:12})
\begin{equation}
    \label{yield}
    0\le y  \le \mathcal{U} \le 100\%,
\end{equation}
where $\mathcal{U}$ refers to the coincidence duty cycle (Table \ref{T1}).

Repeating the above with $\sigma /\mu< 2$ or $0.5$, results vary: the latter is at the edge with S5, S6 and O1, leaving almost no high-quality data. The former slightly increases detector yields by a few percent compared to $\sigma/\mu<1$.

It is remarkable that there is no discernible trend across the different runs, with $y$ of O3a to be similar to that of S5. Fig. \ref{fig:2} presents a summary of this section covering S5 to O3b. Our results illustrate that (\ref{yield}) is $W$-dependent i.e. different search targets will have different detector yield factors.

{{Comparing the $y$-values of each run, presented in the last column of Table. \ref{T2}, with coincidence duty cycles ($\mathcal{U}$) presented in column 4 of Table \ref{T1} and the summary of LVK performance during O2 and O3 in \citep{Davis2021}, it is observed that $y$ is generally less than $\mathcal{U}$ by a factor of about $3-5$. This significant discrepancy is due to two reasons:}}

\begin{itemize}
    \item {Detector yield factors and coincidence duty cycles presented here depend on target of search. We focus on unknown GW transient in the time-frequency domain $W\times B$ with $W$ in \eqref{sub-frame} and $B$ in \eqref{B} representative for the GW signals with the duration similar to that of the descending chirp in GW170817 in the most sensitive frequency band of LIGO. Changing the target of search to signals with shorter duration (e.g., BBH mergers), or longer duration (e.g., signals from pulsars), will change the associated detector yield factor. For instance, seeking for long duration signals with $W=4096$, $y=28\%$ for O2 while $y=9\%$ for O3.}
    
    \item {In \cite{Davis2021} coincidence duty cycles are defined by stable performance of individual detectors in their observing mode. In our approach we look for  simultaneous {\it comparable} nominal performance according to \eqref{ms}. Fig. \ref{overlap} is illustrative for the distinction between these two approaches.}

\end{itemize}

{To provide science quality data, LVK applies vetoes to exclude the data that are seriously affected by noise or glitch \citep{Davis2021}. The criteria for high quality data automatically filters out the moments that the detectors are performing {outside the bounds of \eqref{ms}.}}

\begin{figure*}
\centering
\includegraphics[width=0.95\textwidth]{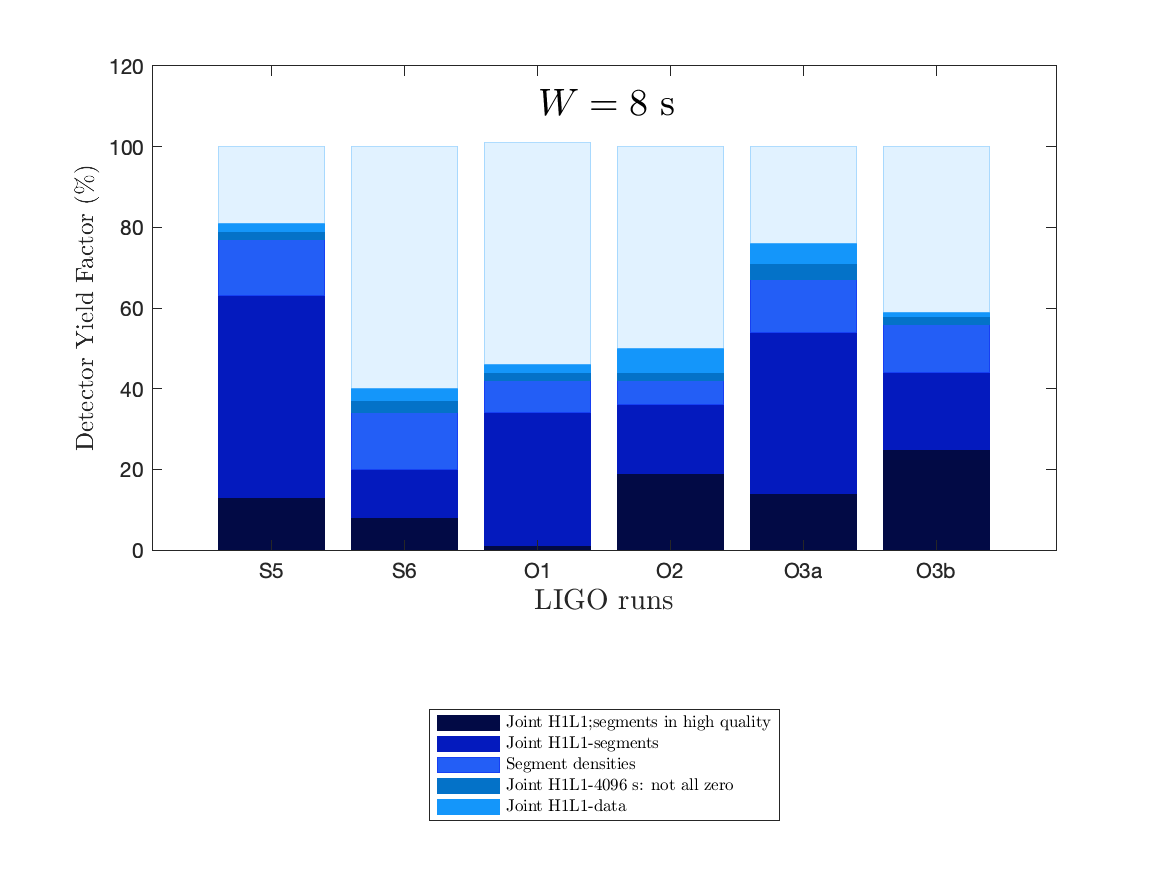}
\caption{\small {Histogram of LIGO performance from S5 to O3b derived from 112 randomly selected frames by GPS time. 
The darkest shade is for high-quality data segments with $\sigma/\mu<1$ and steady phase at the most sensitive frequency range \ref{B} 
{\em in both H1 and L1}. The fraction of segments providing high-quality data is representative for the {\em detector yield factor}. 
O3b features the largest yield factor in search of unknown long duration transients.}
\label{fig:2}}
\end{figure*}

\subsection{Detector yield factors $y$}

The coincidence high-quality observing time of LIGO runs presented here is essential to the conversion of observed-to-true astrophysical event rates. The reported coincidence duty cycle of LIGO detectors spans from $25\%$ to $68\%$ in S5 to O3b and from $40\%$ to $62\%$ in O1 to O3b. The resulting yield factors span from $1\%$ for O1 to $25\%$ for O3b, with no discernible trend across different runs as shown in Fig. \ref{fig:2}.

Fig. \ref{fig:2} summarizes yield factors of the detectors during all LIGO runs, derived from the mean {$\mu$ of the spectrum of each segment (\ref{sub-frame})} for the frequency range (\ref{B}) and the ratios of the standard deviations to the mean for each segment.

Fig. \ref{fig:3} shows that in general, the data in S5 and S6 are widely scattered, expected as these are scientific runs and before aLIGO. For the four later runs, surprisingly, the data for O1, O3a and O3b are scattered, while the data for O2 are placed on a block around the mean value as depicted in Fig. \ref{fig:3}. {
{During} O2 detector performance was substantially more stable than during other runs, although sensitivity {was} less than that of O3 by a factor of about 1.5.} This perhaps explains the paradox that the most exciting event, GW170817, was detected during O2 rather than O3. 

The detector yield factor $y$ is important for true-to-observed event rates of astrophysical transients in EM surveys. 
It evidently affects observing volume 
\begin{equation}
\label{ov}
\left <V T \right >= \left <(\mbox{sensitivity})^3 \times T \times \mbox{coincidence\ duty\ cycle}\right>,
\end{equation}
where $T$ is the total duration of the run. Equation (\ref{ov}) is to be modified by a $y$ to include the true coincidence duty cycle,
\begin{equation}
    \label{TOV}
    \left <V T^{\prime} \right >= \left <V T \right > 
    \times y,
\end{equation}
providing a more precise measure for observational performance of a detector. 
Based on \eqref{TOV}, for the same total observing time, if the sensitivity and detector yield factor go as a ``seesaw'' - sensitivity improves but duty cycle decreases- the total observing volume can remain the same. Therefore, detector performance is not necessarily enhanced even as sensitivity improves. 

Fig. \ref{fig:4} shows true observing volume for the LIGO runs in searches of unknown long duration {transients across segments \eqref{sub-frame}} based on Table \ref{T3}, here normalized to O2. 
Importantly, astrophysical event rate calculations are sensitive to the same $y$ 
relevant, for instance, to heavy element production rates that may further be compared with abundances provided by EM-surveys, 
population statistics of SGRBs and their properties \citep{Kasl2022,Gill2021,Dado2022}. 
Many other astronomical transients provide existing event rates from EM observations (Table \ref{Te}). 

\begin{table*}[]
\begin{tabular}{l||l||l}
\hline
Astrophysical Transient Event & \begin{tabular}[c]{@{}l@{}}Event rate \\ $\mbox{yr}^{-1} \mbox{Gpc}^{-3}$\end{tabular}    & \begin{tabular}[c]{@{}l@{}}Event rate in Local Universe \\ $(100\mbox{Mpc})^{-3} $\end{tabular}                 \\ \hline \hline
Binary Neutron Star (DNS) merger &  $\sim 10-1700$ $^{(1,2,3)}$  &   $\sim 0.1-2\ (\mbox{yr})^{-1} $                                   \\ \hline
Binary Black Holes (BBH) merger  &  $\sim 17.9-44$ $^{(1,2,3)}$  &   $\sim 1\ (\mbox{40yr})^{-1} $                                \\ \hline
Double White Dwarf (DWD) merger  & $\sim 10^{5}$ $^{(4,5)}$   &  $\sim 10^2\ (\mbox{1yr})^{-1} $                                  \\ \hline
SNe Ibc                            &$\sim 10^4$ $^{(6,7,9,10)}$  &   $\sim 10\ (\mbox{1yr})^{-1} $   \\ \hline
Short Gamma-Ray Bursts (SGRB)      & $\sim 1000$ $^{(11)\ \ddagger}$     &    $\sim 1\ (\mbox{yr})^{-1} $                                   \\ \hline
Long Gamma-Ray Burst (LGRB)        & $\sim 1.5\times \mbox{SGRB} ^{\ \ddagger}$       &    $\sim 2\ (\mbox{yr})^{-1} $                          \\ \hline
Core-Collapse SNe                  & $\sim 10^{5}$ $^{(4,12)}$ & $\sim 10^2\ (\mbox{yr})^{-1}$  \\ \hline
Low Luminosity GRB                 & $\sim 100-1800$ $^{(12)}$  &    $\sim 0.1-2\ (\mbox{yr})^{-1} $                                                   \\ \hline
High Luminosity GRB                & $\sim 100-550$ $^{(12)}$    &    $\sim 1-5 (\mbox{10yr})^{-1} $    \\ \hline
Low red-shift Long duration GRB    & $\sim 380$ $^{(6,12)}$  &    $\sim 1\ (\mbox{3yr})^{-1} $       \\ \hline
Super Luminous SNe                 & $ \sim 35-150$ $^{(6,7,8)}$    & $\sim 1-5\ (\mbox{30yr})^{-1} $    \\ \hline
Fast Radio Bursts (FRBs)           & $\sim 10^3-10^5$ $^{(13,14,15,16)}$  &    $\sim 1-10^2\ (\mbox{yr})^{-1} $                                    \\ \hline
X-Ray Flashes                      & $\sim 2$ $^{(17,18)}$     & $\sim 1\ (\mbox{500yr})^{-1} $     \\ \hline
\end{tabular}
\caption{\textit{Estimated event rates of transient events based on EM surveys.}\\
$^{\ddagger}$ \small{GRB rates provided are corrected for beaming \citep{Nakar2020}. Precise measurements of true-to-observed GRB rates due to beaming are not available despite multiple dedicated missions (e. g. BATSE, {\it BeppoSax}, \textit{Swift}, \textit{Fermi}). There appears to be some discrepancy in observed GRB event rates inferred from the BATSE 4B catalog and \textit{Swift}. BATSE suggests LGRBs to be somewhat more numerous compared to SGRBs \citep{Paci1999,Kouv1993}. This seems reversed in {\it Swift} \citep{Wand2010,Wand2015}, although with relatively large uncertainties.}\\
\indent \scriptsize $^{(1)}$\cite{Abbott2021};$^{(2)}$\cite{Abbott2021a} ;$^{(3)}$\cite{Collab2021a} ;$^{(4)}$\cite{Heo2016};$^{(5)}$\cite{Maoz2018} ;$^{(6)}$\cite{GalY2019} ;$^{(7)}$\cite{Li2011}; $^{(8)}$\cite{Froh2021};$^{(9)}$\cite{Dahl2012};$^{(10)}$\cite{Capp2015} ;$^{(11)}$\cite{Nakar2020} ;$^{(12)}$\cite{Guet2007} ;$^{(13)}$\cite{Fial2018} ;$^{(14)}$\cite{Amiri2021};$^{(15)}$\cite{Ravi2019};$^{(16)}$\cite{Bhan2018} ;$^{(17)}$\cite{Yama2002} ;$^{(18)}$\cite{Yama2003}.}
\label{Te}
\end{table*}

\begin{center}
\begin{figure*}[h]
\centering
\includegraphics[scale=0.47]{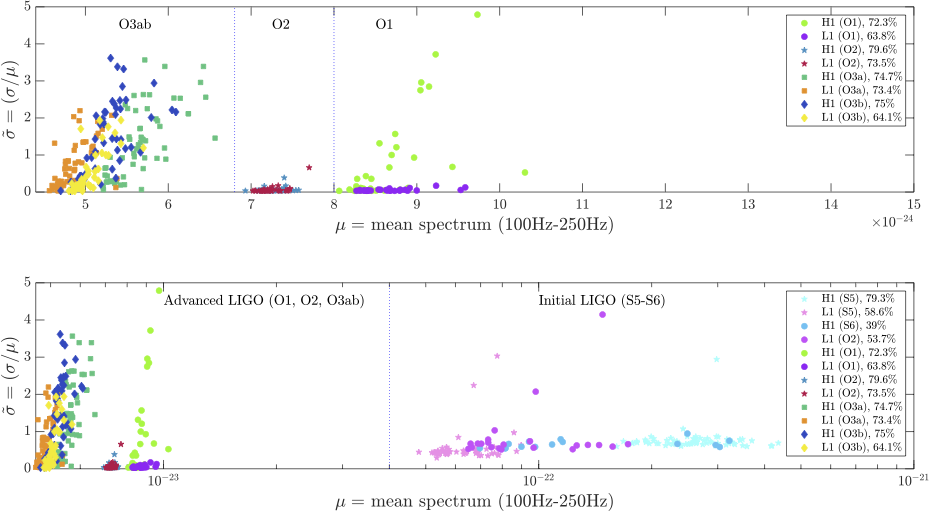}
\caption{\small Detector performance over S5-O3ab in terms of variations($\sigma$) in sensitivity versus sensitivity($\mu$). Shown is a scatter plot of the spectrum($\mu$) vs. for each 4096s data and for the most sensitive frequency range in (\ref{B}). 
Percentages in the legend refer to the fraction of non-zero mean values. There is a significant change from iLIGO to aLIGO (lower panel). Interestingly, O2 shows a highly compact region of small standard deviations and stable performance around the mean. O3a and O3b are similar to each other and despite the improvement in the sensitivity compared to O2, they are more scattered from the mean by having larger $\bar{\sigma}/\bar{\mu}$. 
\label{fig:3}}
\end{figure*}
\end{center}

\begin{figure*}[ht]
\centering
\includegraphics[scale=0.7]{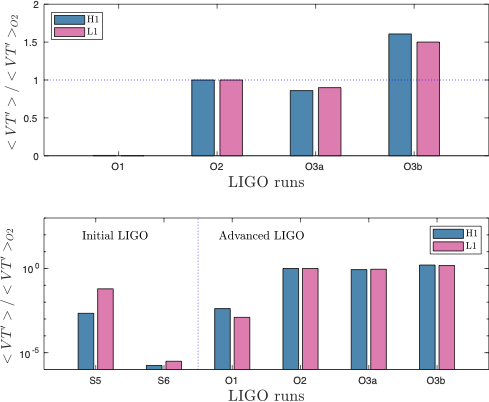}
\caption{\small {Observation volume improvement in LIGO runs from S5 to O3ab, using the inverse of detector noise spectrum representative for the last row of Table \ref{T3}. It shows aLIGO improvement over iLIGO (lower panel), and discovery potential of O3ab similar to O2 (in search of long-duration transients) by true observational volume - essential for detecting unknown GW signals and event rate estimation (upper panel).}} 
\label{fig:4}
\end{figure*}

\begin{table*}[]
\centering
\footnotesize
\begin{tabular}{|c|c|c|c|||c|c|c|c|}
\hline
Run                                                           &                     & S5                                     & S6                                     & O1                                     & O2                                     & O3a                                    & O3b                                   \\ \hline \hline
\multirow{4}{*}{Sensitivity ($1/\sqrt{Hz}$)}                  & \multirow{2}{*}{H1} & \multirow{2}{*}{\pmb{$1.02 \times 10^{-21}$}} & \multirow{2}{*}{\pmb{$7.27 \times 10^{-22}$}} & \multirow{2}{*}{\pmb{$1.037 \times 10^{-23}$}} & \multirow{2}{*}{\pmb{$7.3 \times 10^{-24}$}} & \multirow{2}{*}{\pmb{$5.23 \times 10^{-24}$}} & \multirow{2}{*}{\pmb{$4.86 \times 10^{-24}$}}  \\
                                                              &                     &                                        &                                        &                                        &                                        &                                        &                                       \\ \cline{2-8} 
     & \multirow{2}{*}{L1} & \multirow{2}{*}{\pmb{$3.3 \times 10^{-22}$}} & \multirow{2}{*}{\pmb{$5.9 \times 10^{-22}$}} & \multirow{2}{*}{\pmb{$1.52 \times 10^{-23}$}} & \multirow{2}{*}{\pmb{$7.2 \times 10^{-24}$}} & \multirow{2}{*}{\pmb{$5.07 \times 10^{-24}$}} & \multirow{2}{*}{\pmb{$4.88 \times 10^{-24}$}}  \\
                                                              &                     &                                        &                                        &                                        &                                        &                                        &                                       \\ \hline
\multirow{2}{*}{Detector Yield ($y_D$)(\%)}  & \multirow{2}{*}{}   & \multirow{2}{*}{{\bf 13}}                   & \multirow{2}{*}{{\bf 8}}                   & \multirow{2}{*}{{\bf 1}}                  & \multirow{2}{*}{{\bf 18.7}}                  & \multirow{2}{*}{{\bf 13.6}}                   & \multirow{2}{*}{{\bf 25.3}}                  \\
                                                              &                     &                                        &                                        &                                        &                                        &                                        &                                       \\ \hline
\multirow{4}{*}{$VT^{\prime} = VT\times y$}  & \multirow{2}{*}{H1} & \multirow{2}{*}{\pmb{$2.6 \times 10^{-7}$}}  & \multirow{2}{*}{\pmb{$2.09 \times 10^{-10}$}}  & \multirow{2}{*}{\pmb{$5 \times 10^{-7}$}}  & \multirow{2}{*}{\pmb{$1.2 \times 10^{-4}$}}  & \multirow{2}{*}{\pmb{$1.03 \times 10^{-4}$}}  & \multirow{2}{*}{\pmb{$1.9 \times 10^{-4}$}} \\
                                                              &                     &                                        &                                        &                                        &                                        &                                        &                                       \\ \cline{2-8} 
    & \multirow{2}{*}{L1} & \multirow{2}{*}{\pmb{$7.9 \times 10^{-6}$}}  & \multirow{2}{*}{\pmb{$3.9 \times 10^{-10}$}}  & \multirow{2}{*}{\pmb{$1.6 \times 10^{-7}$}}  & \multirow{2}{*}{\pmb{$1.3 \times 10^{-4}$}}  & \multirow{2}{*}{\pmb{$1.14\times 10^{-4}$}}  & \multirow{2}{*}{\pmb{$1.9 \times 10^{-5}$}} \\
      ($\mbox{Gpc}^3 \mbox{yr}$)    &                     &                                        &                                        &                                        &                                        &                                        &                                       \\ \hline
\multirow{4}{*}{$VT^{\prime}$/$(VT^{\prime})_{O2}$} & \multirow{2}{*}{H1} & \multirow{2}{*}{\pmb{$2.2 \times 10^{-3}$}}             & \multirow{2}{*}{\pmb{$1.7 \times 10^{-6}$}}    & \multirow{2}{*}{\pmb{$4.2 \times 10^{-3}$}}                & \multirow{2}{*}{{\bf 1}}                     & \multirow{2}{*}{\pmb{$0.86 $}} & \multirow{2}{*}{\pmb{$1.61$}}                \\
     &                     &                                        &                                        &                                        &                                        &                                        &                                       \\ \cline{2-8} 
    & \multirow{2}{*}{L1} & \multirow{2}{*}{\pmb{$0.620$}}    & \multirow{2}{*}{\pmb{$3.07 \times 10^{-6}$}}             & \multirow{2}{*}{\pmb{$1.2 \times 10^{-3}$}}                & \multirow{2}{*}{\pmb{1}}                     & \multirow{2}{*}{\pmb{$0.90$}}                & \multirow{2}{*}{\pmb{$1.50$}}                \\
    &        &     &     &       &  &     &   \\ \hline
\end{tabular}
\caption{\small {\textit{Summary of LIGO detector performance in terms of observing volume $\left <VT^{\prime} \right >$}. LIGO detectors observational volume improvements in the most sensitive frequency range ($100-250 \mbox{Hz}$) during the 6 runs S5-O3b. Observational volume is defined as \eqref{TOV}, where $V$ is the spatial volume defined by the sensitivity of the detector in each run, $T$ is the observing time and $T^{\prime}$ is the $y$-corrected observational time considering the dimensionless detector yield factor $y$ in each run. 
The last row provides the $<VT> \times \mbox{yield}$ values normalized to O2 as it was the host for the only multi-messenger LIGO event GW170817. To be more clear, S5 and S6 was scientific runs during iLIGO, but the other four are observational runs in aLIGO. The last two rows show that O2 and O3ab are almost on par with each other as they have similar observing volume.} }
\label{T3}
\end{table*}

As per Table \ref{Te}, EM-surveys offer comprehensive yet somewhat imprecise statistics of our transients of interest. This is apparent in considerable uncertainties in event rates that appear to be statistic and/or systematic. Future joint EM-GW surveys may come to the rescue. This will pose additional challenges (if not opportunities) for consistent event rates. Inevitably, this calls for accurate and high detector yield factors $y$.

GW170817 serves as a case at hand. Detected in O2 at a distance of $40 \mbox{Mpc}$, it represents a DNS coalescence followed by an SGRB, X-ray, UV, optical and radio-emissions. This was the only event of the type within about 269 days of O2. During O2 the horizon distance of LIGO was about $100 \mbox{Mpc}$ with a coincidence duty cycle of $45\%$. Assuming GW170817 is representative of typical DNS merger, these numbers give a rough estimate of $3 \mbox{yr}^{-1} (\mbox{100Mpc})^{-3}$ for DNS coalescences. Including accurate detector yield factors $y$, the event rate increases by a factor of about three. If all SGRBs to be counterparts of DNS mergers, this affects the estimated event rates for SGRBs and kilonovae associated with DNS mergers. A comparison with EM-surveys shows the resulting event rates to have some tension, though inconclusively due to the $n=1$ event count with GW170817 in O2 and the somewhat uncertain estimates from EM-surveys. {Further comments are deferred to \S\ref{Sec:Co}.}

\section{LIGO Improvement Factors} \label{Sec:IF}
Our results in the previous section point toward appreciable room to enhance discovery power, some of which may be realizable in advance of much anticipated next generation detectors \citep{Abbott2017b,Abbott2017e,Torr2019}. 
Alternatively, next generation detectors, such as the Cosmic Explorer and Einstein Telescope, are anticipated to have a horizon that covers high redshift cosmological distances (at least an order of magnitude larger than aLIGO), essential for detecting numerous events \citep{Dwye2015,Digg2017}. Consequently, they promise to provide precise measurements of transients event rates.

Focusing on O4-5, we highlight some factors to enhance the discovery potential of our current generation of detectors.

\subsection{{Unrealized Sensitivity}} \label{ss1}

As one of the most important factors in detection performance and discovery potential, sensitivity has been always been a primary objective in each new run. From S5 to O3b, detector sensitivity has seen improvements by over two orders of magnitude. Further improvements are expected by a factor of about 2-2.5 in O5 about 2027 \citep{Abbott2020}. 
This is expected to result in an additional increase of discovery power by a factor of about 10 for observational times similar to existing runs.

\subsection{Detector yield factor in EM-GW surveys} \label{ss2}

{Our focus is on true duty cycle ($y$), defined by a detector yield factors key to discovery power of new and unknown signals.
Currently, $y$-values are optimistically about $10-20\%$}  -  considerably below that of conventional EM-observatories \citep[e.g.][]{nasac}. 
{It also challenges searches for MM-signals by missed events when continuous coverage is poor.}
In particular, this would apply to transient events of interest, notably SGRBs associated with DNS mergers and GW-emission from central engines of CC-SNe. According to EM-observations, the event rate of SGRBs is $\mbox{1000 yr}^{-1}\mbox{Gpc}^{-3}$ \citep{Nakar2020}. 
{If all SGRBs derive from {DNS mergers are similar to GW170817 and} $y\simeq 20\%$ (similar to O2-O3), it might take some a half-century to see another GW170817-like event at similar proximity - or more if $y\simeq 10\%$ (Table \ref{T2}). Even though $y\simeq 100\%$ may not be realistic, 
yields similar or better than O2/O3ab seem desirable.
High detector yield factors seems pertinent to discovery of the unknown in joint EM-GW surveys.}

\subsection{Time-symmetric data analysis}\label{ss3}

{Data analysis methods are essential to realizing discovery power, even for resolving known events for their detailed physical and astronomical aspects. Quite generally, astrophysical transient sources tend to produce ascending GW-chirps, as in CBC, or descending GW-chirps, in spin-down of a compact central engine powering a GRB or CC-SN.
In searching for the unknown, an optimal analysis method should be equally sensitive to these two classes of signals: the method of analysis should be {\em time-symmetric}.}

In contrast, LIGO currently employs highly optimized methods to {detect CBC \citep{Allen2012}, notably GW170817 with ${\cal E}_{GW}\simeq 2.5\% M_{\astrosun}c^2$ by essentially ideal matched filtering against known wave-forms. LIGO has also multiple pipelines for searching for unknown signals for instance cWB \citep{Drago2021}, Hidden Markov Model \citep{Suvo2016} and Bayeswave \citep{Corn2015}, among others.} Yet, a LIGO threshold of about ${\cal E}_{th,GW}\geq 650\%M_{\astrosun}c^2$ is reported for unknown GW-signals from a putative long-lived compact remnant \citep{Sun2019}; see further Fig. 1 of \citep{Abbott2017e}. Over two times the total mass-energy of the DNS GW170817, this threshold violates (\ref{EC1}).

Fig. \ref{fig:11} shows search domains for descending signals defined by (\ref{EC1}), indicating {\it ``Allowed''} and {\it ``Excluded''} zones. 
LIGO explorations to-date for un-known GW-signals have been performed mostly in the Excluded zone, leaving most of the Allowed region {\it ``Uncharted''}. In fact, for long-lived remnants, the entire Allowed region remains Uncharted 
{by the above mentioned LIGO search algorithms} \citep{Abbott2019a}.

When ${\cal E}_{th,GW}\gtrsim 650\%M_{\astrosun}c^2$, (\ref{EC1}) is not satisfied. Furthermore, it is ostensibly unbalanced, limiting sensitivity to an un-modeled descending signal to about $0.3\%$ of a preceding merger measured by energy output ${\cal E}_{GW}$. By amplitude, the corresponding discrepancy is a factor of 17 between ascending merger chirps and any potential descending signal from spin-down of a compact remnant.

A time-symmetric method of analysis principally uses some form of matched filtering over a bank of time-symmetric templates. Notable examples are time-sliced Fourier-based methods and butterfly matched filtering \citep{Putten2019,Putten2019a}.

To illustrate time-symmetry, we apply conventional time-sliced {Fast Fourier Transform (FFT) \citep{Fourier1822,Gauss1805,cool1965} to GW170817}. GW170817 is presently the only event to date of interest by the associated GRB170817A with contemporaneous emission in a descending GW-chirp, recently identified with spin-down of a Kerr black hole \citep{Putten2022}.
Fig. \ref{fig:TR}, clearly confirms the merger chirp from $\sim 100 \mbox{Hz}$ to $\sim 300$, 
representing ${\cal E}_{GW}\simeq 2.25\% M_{\astrosun}c^2$ output in gravitational radiation {($2.5\%M_\odot c^2$ in a run-up to 650\,Hz \citep{Abbott2017f})}. 
The result is evidently invariant by doing the same calculation following a time reversal. Importantly, the signal is detectable in \textit{both} detectors. The glitch in L1 during the merger is perhaps illustrative for the need of having (at least) two detectors working at the same time producing high-quality data. 

Interestingly, Fig. \ref{fig:5} shows a faint yet discernible trace { of the above-mentioned {(descending)} GW-signature of GRB170817A in the time-sliced FFT spectrogram. 
{Notably, we have consistency in representing a total output of  ${\cal E}_{GW}\simeq 3.5\% M_{\astrosun}c^2$ for the descending chirp with a more detailed analysis of the descending chirp of black hole spin-down, recently confirmed at the 5.5$\sigma$ confidence level  (Fig. \ref{AA} and \cite{Putten2022}.}
Here, the latter is {also} seen to be similar noting ${\cal E}_{GW}\simeq 2.25M_\odot c^2$ of the merger chirp at slightly lower frequencies, where LIGO sensitivity is relatively high. 

{The results of these two independent approaches reveal a descending GW-chirp during GRB170817A at $\mathcal{E}_{GW}\simeq 3.5\%M_{\odot} c^2$:} {\em the Uncharted region is not empty.} Therefore, \eqref{EC1} needs to be satisfied by a detection threshold of order $1\% M_{\odot}c^2$, far below the threshold $E_{GW} = 650\%M_{\odot}c^2$ of aforementioned power-excess methods \citep{Sun2019}.

\subsection{{Improving discovery power}}\label{ss4}
Following \S\ref{ss1}, \S\ref{ss2} and \S\ref{ss3} we estimate potential room for an improvement in discovery power to the unknown in the current generation of LIGO detectors, calculated by the four-dimensional volume $\left <VT\right>$ over O4-5. Compared to O3, we estimate:

\begin{equation}
\label{improvement1}
\dfrac{(VT^{\prime})_f}{(VT^{\prime})^{LIGO}_{today}}=\mathcal{A}\times \mathcal{B} \sim \mathcal{O}\left( 10^5 \right),\vspace{0.2cm}
\end{equation}

where $\mathcal{A}$ and $\mathcal{B}$ are representative for potential improvements in LIGO hardware and, respectively, software, 
\onecolumngrid
 \begin{eqnarray}
 \begin{array}{lll}
     \label{improvement2}
     &&\mathcal{A}\simeq \left( \dfrac{h_{O5}}{h_{On}}\right)^3 \times\left(\dfrac{100\%}{Y_{On}}\right)\sim \left\{ \begin{aligned} 
  \mathcal{O}\left( 10^4 \right) \hspace{0.2cm} n=1\\
  \mathcal{O}\left( 10^2 \right) \hspace{0.2cm} n=2\\
  \mathcal{O}\left( 10^2 \right) \hspace{0.2cm} n=3
\end{aligned} \right.  \\ \\ 
     &&\mathcal{B}\simeq \left( \dfrac{\mbox{LIGO sensitivity to CBC}}{\mbox{LIGO sensitivity to descending signals}}\right)^{3/2} \sim \mathcal{O}\left( 10^3 \right).
 \end{array}
 \end{eqnarray}
 \twocolumngrid
 Here, ${\cal A}$ is derived from potential improvements in the detector yield factor; ${\cal B}$ is derived from the above-mentioned ${\cal E}_{GW}\simeq 2.5\% M_{\astrosun}c^2$ in GW170817 and a reported threshold of ${\cal E}_{GW}={\rm few}\times M_{\astrosun}c^2$ \citep{Abbott2017e,Sun2019}. Bringing down the latter to a few percent {\em on par with CBC} by time-symmetric data-{analysis shows ${\cal B}\simeq 4200$.} Estimating $\mathcal{B}$ during O3 gives the same result as the same analysis methods were employed compared to O2 (Fig. 2 and Table I in \cite{Abbott2021b}).

 Up to an order of magnitude, $\mathcal{A}$ and $\mathcal{B}$ each point to considerable expansion of discovery power in the coming years.
 Even if hardware and software improve at a fraction of their limits, considerable improvements in discovery power are expected. 
 For instance,  $\mathcal{B}$ combined with joint H1 and L1 duty cycles for quality data at about $60\%$ presents an outlook on discovery power (\ref{improvement1}) improving by about {393000} in the planned O5 run, compared to O3. On this basis, {present observations appear to be just the tip of the iceberg}.
 
\begin{figure*}
\centering
\includegraphics[scale=0.45]{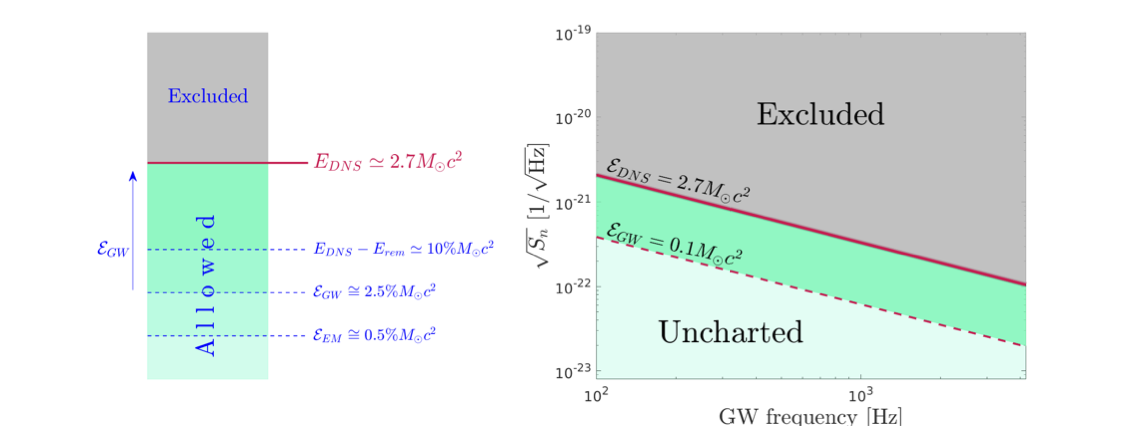}
\caption{\small The search domains for a GW170817 descending signal according to \cite{Abbott2017e}. 
(Left.) Radiation energies emitted following a merger must satisfy energy conservation \eqref{EC1}. 
Specialized to GW170817, $E_P$ is given by the total mass-energy $E_{DNS}\simeq2.7M_{\astrosun}c^2$ is the total mass-energy of the progenitor and \eqref{EC1} defines a required energy threshold of detection $\mathcal{E}_{th}\leqq E_{DNS}$: $\mathcal{E}>E_{DNS}$ represents an {\it ``Excluded zone''} and $\mathcal{E}<E_{DNS}$ a zone of {\it ``Allowed''} radiation energies. 
Possibly, the merger output $\mathcal{E}_{GW}\simeq 2.5\%M_{\astrosun}c^2$ sets a scale for potential descending chirps in GW-emission,
further suggesting a reduced search domain of interest $\leq 10\%M_{\astrosun}c^2$. Notably, EM-radiation is only $\mathcal{E}_{EM}\simeq0.5\%M_{\astrosun}c^2$ \citep{Mooley2017,Mooley2018}. (Right.) The Excluded (grey) and Allowed (green) zones are shown in the frequency-noise amplitude spectrum of LIGO during O2, based on Fig. 1 in \cite{Abbott2017e}. High-ligthed is the {\it ``Uncharted zone''} covering most of the Allowed zone, not explored by any of the LIGO searches. Notably, LIGO searches for long-lived remnants are entirely in the Excluded zone in light of a search threshold $\mathcal{E}_{th,GW}\geq6.5M_{\astrosun}c^2$ \citep{Sun2019,Abbott2019a}, {explaining the non-detection in these earlier post-merger searches.}}
\label{fig:11}
\end{figure*}

\begin{figure*}
\centering
\includegraphics[scale=0.5]{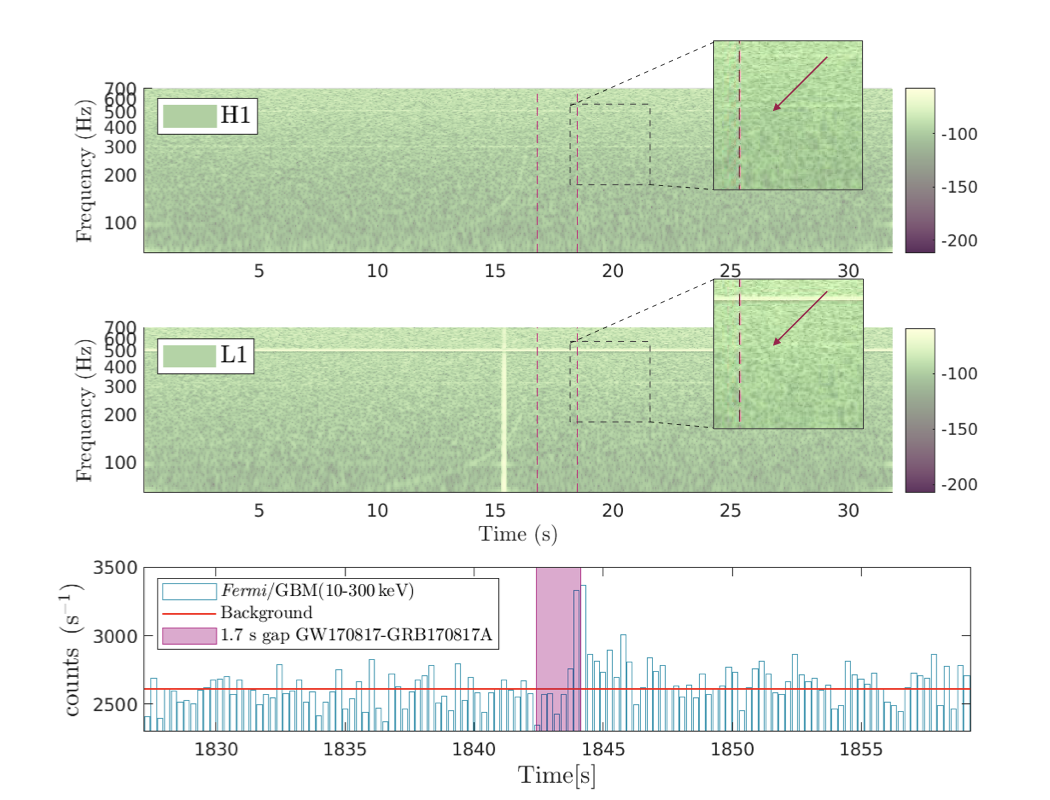}
\caption{\small The upper and middle panels show spectrograms of 32\,s of data around GW170817 for H1 and L1 by {\sc GWXplore}. Discernible though faint is a descending chirp following the merger. 
{This benchmark, together with Fig. \ref{fig:TR}, is illustrative for a time-symmetric exploration of the Uncharted region in Fig. \ref{fig:11}}. The lower panel shows the time delay time of 1.7\,s between the GW170817 trigger and the time-of-onset of GRB170817A {({\it Fermi}-GBM panel after \cite{Putten2022}}, and the moment that the descending chirp appears visible. Although the exact starting point of the descending signal is not well resolved in the present spectrogram, it appears contemporaneous with GRB170817A attributed to MM-emission from a black hole disk or torus system for the lifetime of black hole-spin \citep{Putten2022}.
Notably, the lifetime of the primary HMNS {is} of about 0.92\,s, when it experiences delayed gravitational collapse to a Kerr black hole - in the 1.7\,s gap between GW170817 and GRB170817A \citep[see also ][]{Murg2021}.}
\label{fig:5}
\end{figure*}

\begin{figure*}
\centering
\includegraphics[width=1\textwidth,trim=2 2 2 2,clip]{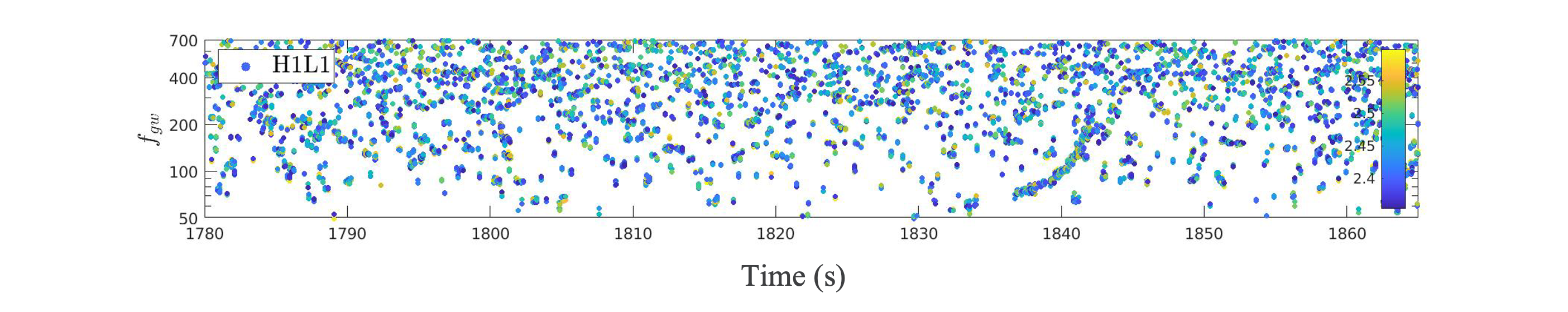}
\caption{\small 
{The descending GW-chirp during GRB170817A observed at high contrast in a merged (H1,L1)-spectrogram, produced by butterfly matched filtering. This descending signal starts about 0.92\,s post-merger, covering the frequency range $700\sim 200$Hz. It is confirmed to be associated with the spin-down of the central engine of GRB170817A with a confidence level of $5.5\sigma$ (Reprinted from \citealt{Putten2022}).}
}
\label{AA}
\end{figure*}

\section{{Diversity in DNS and MM-afterglows}} \label{Sec:DNS}
Neutron stars are the remnant of the majority of massive stars producing CC-SNe. In general, the lifetime of a (proto-)neutron star depends on its mass \citep[e.g.][]{Lucca2020}. Hyper-massive neutron stars (HMNS) that may be produced in the immediate aftermath of a DNS merger such as GW170817 tend to have a relatively short lifetime \citep{Lipp2017} before experiencing gravitational collapse to a black hole, while supermassive neutron stars (SMNS) may live forever. 

DNS systems are interesting as progenitors of MM-astronomical transients. {GW170817 produced the SGRB 
(GRB170817A) and kilonova 
(AT2017gfo). Normal long GRBs, on the other hand,} are believed to originate from CC-SNe \citep{Troj2017,Eich1989,Rezz2011,Nakar2006,Leva2016,Berg2014}. {A DNS merger sequence can be complex, depending on {total mass of the system}, producing an SMNS or HMNS, the latter with prompt or delayed collapse to a black hole. As the {total mass of the system} given by the two neutron star progenitors is key factor to the final outcome, we here first recall the mass distribution of neutron stars provided by existing radio-surveys.}

{
{Radio surveys reveal a bimodal distribution for both neutron stars in DNS systems, and total mass of the binary system \citep{Oeze2012,Schw2010}. Neutron stars masses range from about $1 M_{\astrosun}$ to slightly more than $2M_{\astrosun}$. The available EM-data derives from neutron stars associated with another neutron star or white dwarf, also from optical binaries \citep{Oeze2012,Taur2017,Roch2019,Lina2020}.}

{As DNS mergers are of particular interest, we will continue with only these included in radio-selected surveys. Neutron stars in DNS systems usually consist of a recycled fast spinning neutron star with a slowly spinning companion neutron star. 
These two populations were reported to have distinct mass distributions, the first with a bimodal mass-distribution, the second with a uniform mass-distribution \citep{Farr2019}.}

{DNS systems are often classified in two categories, according to merger times shorter or longer than the Hubble time \citep{Andr2019}. This can be inferred from the relation between the total mass of a DNS system and its orbital period by orbital separation. Fig. \ref{fig:9} shows the statistics of the known DNS systems,  showing no further dependency between the orbital separation and mean mass or, equivalently, lifetime of the progenitor DNS system. Intrinsically, these two categories are on the same footing. }

{Considering all of the neutron stars in DNS systems with precisely measured components and total mass \citep{Taur2017,Seng2022,Stov2018,Mart2017,Lync2018,Came2018}, the 
mass-distribution of DNS components and merger remnants is plotted in Fig. \ref{fig:7}. Excluded is a $10\%$ energy loss of the system for the remnant mass-distribution, representative for gravitational radiation and mass loss {in the process of coalescence}. Both distributions are representative for a normal mass-distribution function \citep{Lina2020}, consistent with the first 
{peaks in Fig. \ref{fig:7}}. }

{While DNS systems are represented by the first peak, the second peak corresponds to neutron stars in a binary with a white dwarf or X-ray binaries 
{, not included in this study. Fig. \ref{fig:7} suggests} that a DNS merger of two neutron stars at the left tail of the first distribution can form an SMNS belonging to the same mass-distribution with a long lifetime and possibly so without any major afterglow. } 

\begin{figure}
\centering
\includegraphics[scale=0.37]{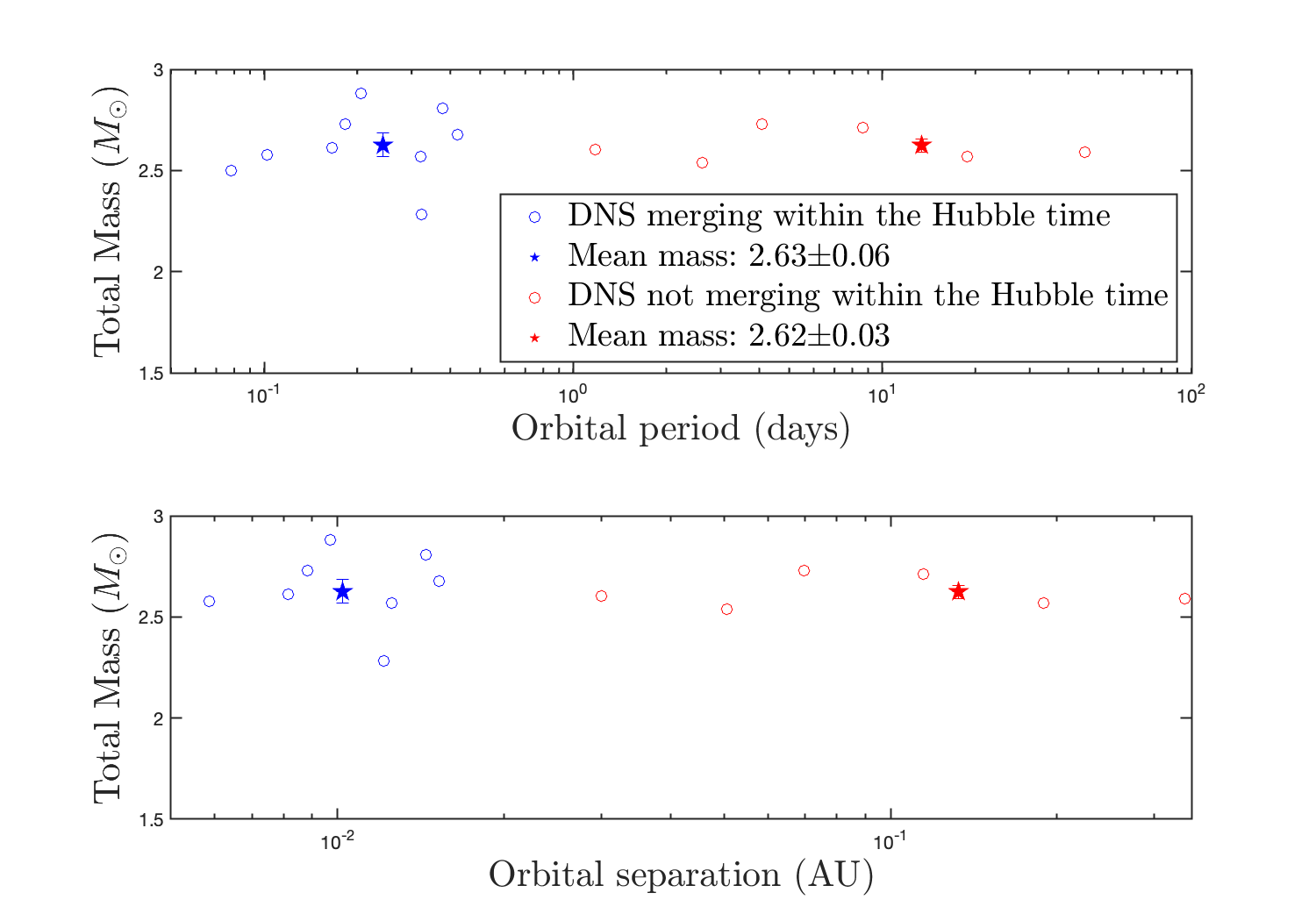}
\caption{\small A scatter plot of total mass vs. orbital period (top panel) and orbital separation (lower panel) of known DNS systems \cite{Seng2022,Taur2017}. Results show no distinction between DNS systems merging within or longer than a Hubble time.\label{fig:9}}
\end{figure}

\begin{figure*}
\centering
\includegraphics[scale=0.5]{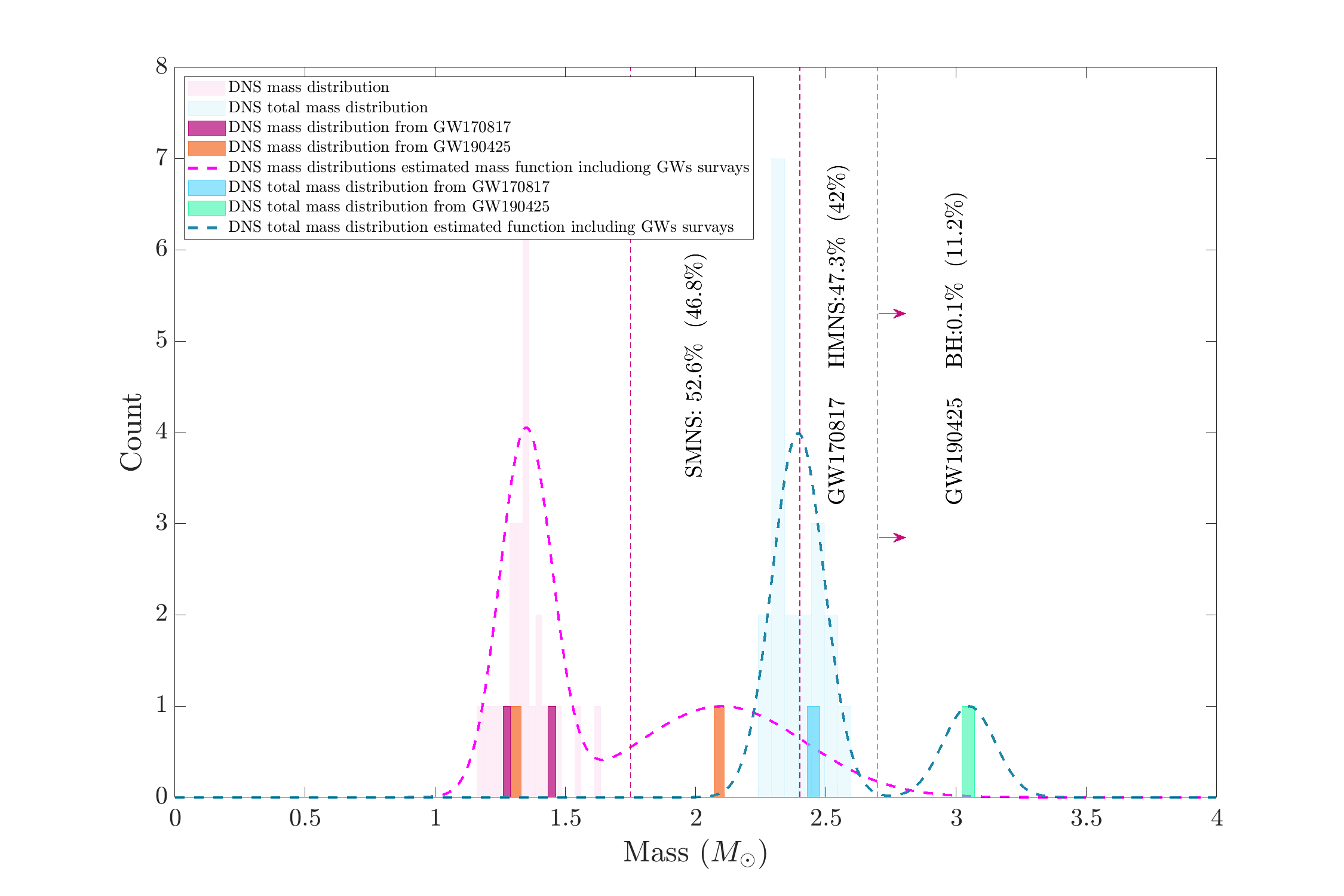}
\caption{\small 
{Mass-distribution of NS in DNS systems from radio-surveys and GWs and their merger remnant. The pink curve is for mass-components while the blue curve shows merger remnant mass {($10\%$ of their total mass is to account for GW-emission and mass-loss)}. The second peak in pink curve represents the mass distribution of NS in binaries with a WD or in X-ray binaries, which are not shown here.} Considering both populations, and $2.4\mbox{M}_{\astrosun}$ and $2.7\mbox{M}_{\astrosun}$ as a {putative} lower and upper threshold for HMNS masses, $47\%$ of DNS mergers 
{are expected to produce an} SMNS while $42\%$ are expected to
produce an HMNS with a relatively short lifetime as in GW170817. 
{A remaining $11\%$ of the DNS mergers are expected to experience 
prompt gravitational collapse to a black hole.}}
\label{fig:7}
\end{figure*}

{Marked in different colors in Fig. \ref{fig:7}, we provide a first
GW-survey of DNS systems consisting of GW170817 ($M_1=1.46M_{\astrosun},\ M_2=1.27M_{\astrosun},\ M_T=2.73M_{\astrosun}$) and GW190425 ($M_1=2.1M_{\astrosun},\ M_2=1.3M_{\astrosun},\ M_T=3.4M_{\astrosun}$) \citep{Abbott2019,Collab2021a}.} Shown is a significant difference in the distribution pattern. Apart from a slight change in the first peak, GW190425 seems to support a bimodal distribution of DNS systems and remnant mass, while GW170817 follows the original radio-survey based mass-distribution. 

\subsection{{GW170817 and GW190425}}
While GW170817 and GW190425 are both derived from DNS mergers, they are nevertheless distinct. 

GW170817 has an associated SGRB GRB170817A observed 1.7\,s after the merger \citep{Gold2017}. Within 100 days it featured emission in X-rays, optical and radio seen by INTEGRAL, \textit{Fermi}-GBM, the Hubble Space Telescope (HST), the Chandra X-ray Observatory and the Karl G. Jansky Very Large Array \citep{Abbott2017d,Troj2017,Hall2017,Abbott2017g,Gold2017,Marg2018}. 
{These MM-observations provide strong evidence for the GW170817-GRB170817A association. The kilonova AT2017gfo reveals a further production of heavy elements \citep{Gill2021}.
The central engine of GRB170817A is identified with a Kerr black hole through a contemporaneous descending chirp in gravitational radiation, attributed to delayed gravitational collapse of the initial HMNS \citep{Putten2019a,Putten2022}.} 

{GW190425 presented a second DNS merger candidate. Searches for GW170817-like MM-afterglow emission produced no confirmed results. This may, in part, be due to the large (imprecise) localization area, a significantly larger distance compared to GW170817 and a relatively heavy remnant with possibly prompt gravitational collapse {to a black hole \citep{Leva2020,Anti2020,Coug2019a}.}
GW190425 is detected by L1 when H1 was not in observation mode \citep{Collab2020}.
While GW190425 is subject to considerable observational uncertainties, it nevertheless appears to be distinct from GW170817 by no detection of MM-afterglow emission and relatively high total mass.}

{As mentioned, the second peak in the first bimodal distribution (pink) in Fig. \ref{fig:7} shows NS in WD-NS binaries}. GW190425 appears to represent the merger of a low mass companion $M_2=1.3 M_{\astrosun}$ from the first with a massive member of this second peak. 
The sum is possibly sufficiently massive to result in prompt gravitational collapse of the initial HMNS, leaving negligible window for an accompanying kilonova, while a GRB would unlikely be observable due to beaming.}

{From Fig. \ref{fig:7}, we can infer the lifetime of the initial remnant of DNS coalescence, arbitrarily long for an SMNS and finite for an HMNS before gravitational-collapse to a black hole. 
{Fig. \ref{fig:7} suggests} SMNS below $2.4 M_{\astrosun}$ are relatively long-lived. Heavier yet below $2.7-3 M_{\astrosun}$, HMNS are formed with finite but short lifetimes. 
{The lifetime of more massive HMNS is expected to be essentially zero,} 
marking prompt gravitational-collapse to a BH 
in the immediate aftermath of the merger. 
These possible regimes are highlighted with dashed lines in Fig.\,\ref{fig:7}.}


\begin{figure}
\vskip0.0in
\centering\includegraphics[scale=0.6]{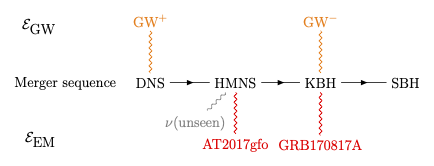}
\caption{\small {\textit{Merger sequence of a GW170817-like event}. The merger signal (GW$^+$) terminates in the formation of the initial HMNS powering the kilonova AT2017gfo. In delayed gravitational collapse, it produces a Kerr black hole powering GRB170817A, signaled by a descending gravitational wave signal (GW$^-$) \citep{Putten2022}.}
\label{fig:009}}
\end{figure}
\begin{figure}
\vskip0in\centering\includegraphics[scale=0.5]{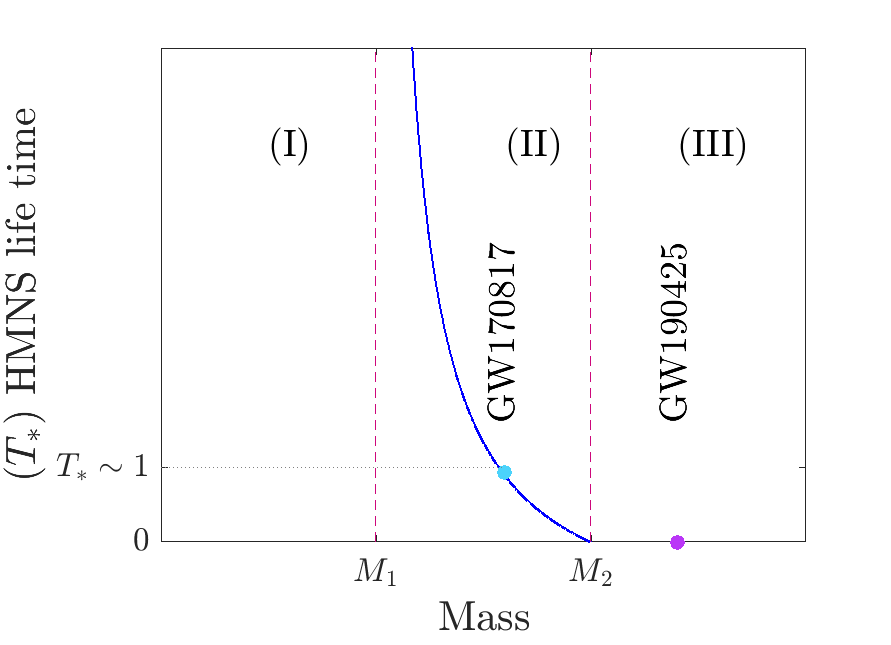}
\caption{\small Schematic of the lifetime of an HMNS as a function of total mass of the progenitor DNS merger. {Here, $M_1$ refers to the cutoff at the left tail of the blue curve in Fig.\,\ref{fig:7}, while $M_2$ refers to the cutoff at the right tail of the same curve.}
\label{fig:09}}
\end{figure}

{The expected dependence of the lifetime of the initial remnant HMNS on total mass of the binary system \citep[e.g.][]{Lucca2020} is consistent with GW170817 and GW190425. Specifically,
 GW170817 has a total mass is $M_T=2.73 M_{\astrosun}$ 
 ($M_T=2.45 M_{\astrosun}$ reduced by $10\%$ due to losses in gravitational-radiation and mass ejecta), leaving
 an initial HMNS remnant with relatively short but non-zero lifetime. 
 Based on the kilonova AT2017gfa, the estimated lifetime of the HMNS is
 $t_w\left(0.98\pm 0.3\right)$\,s \citep{Gill2019},
consistent with the time of onset $t_{s,GW}$ of a descending chirp in GW-emission from a black hole-disk or torus system following gravitational-collapse \cite{Putten2022},
\begin{equation}
\label{time}
t_{s,GW}-t_m \simeq (0.92 \pm 0.08)\ {\rm s},
\end{equation}
where $t_m$ is the merger time.   
(\ref{time}) improves on a previous estimate of the same \cite{Putten2018,Putten2019a} reviewed in \cite{Murg2021}.
In contrast, GW190425 has a total mass of $M_T=3.4 M_{\astrosun}$
in the second peak (blue curve, Fig.\,\ref{fig:7}) and above the 
threshold of $3M_{\astrosun}$, even after the reduction of $10\%$. 
Prompt gravitational collapse of its initial remnant naturally explains the non-detection of a kilonova, while the non-detection of a GRB may be attributed to beaming. A black hole torus system may have formed with weak kilonova by the relatively low MeV-temperatures of the torus compared to the temperature of the HMNS in GW170817 (further below) with accompanying SGRB or SGRBEE depending on black hole-spin \citep{Putten2001} - GW190425 is too far out to probe for a contemporaneous descending chirp in gravitational radiation.}

\subsection{{Kilonovae from HMNS or BH-torus systems?}}

{Conceivably, kilonovae may derive from MeV-neutrino driven winds from a newly formed massive NS \citep[cf.][]{Gill2019} or, following its collapse, from a disk or torus around a newly formed black hole.} The difference, if any, is related to duration and temperature. Torus-driven kilonovae driven by neutrino-winds at $T_D\simeq 2$\, MeV are powered by outflows of about $\dot{m}\simeq 0.05\%M_\odot$s$^{-1}$ {(see Fig 10 of \cite{Putten2003}). Torus-driven kilonovae may} appear moderate compared to those driven by neutrino-winds at temperatures $\gg10$\,MeV of an HMNS newly formed (cf. MeV-neutrino light curve of SN1987A) in the immediate aftermath of a merger. The significance to the r-process in kilonovae may be comparable if such torus-wind is sufficiently long-lived.

GRB211211A might be illustrative, upon identifying its duration $T_{90}\sim 50$\ s with the lifetime of a black hole-torus system - post-merger to a DNS or BH-NS merger. A kilonova derived from a mass-ejecta
\begin{eqnarray}
m_{ej}=\dot{m}T_{90} \simeq 2.5 \%M_{\odot}
\label{kn11A}
\end{eqnarray}
will be on par with AT2017gfo, derived from the HMNS produced in GW170817A over its relatively short lifetime of $\lesssim 1$\ s \citep{Marg2017}.

Torus-driven kilonovae may be challenging to detect at the distance of GW190425, however. This scenario appears consistent with numerical simulations for GW190425 with no definite EM counterpart \citep{Diet2019}.

The above is based solely on the mass-distribution of DNS remnants in Fig.\,\ref{fig:7}, $46\%$, allowing the formation of an SMNS from the left tail of the distribution (pink curve). $42\%$ are GW170817-like, producing an HMNS with finite lifetime with kilonova output and, subsequent to delayed collapse to a black hole surrounded by a finite amount of merger debris, an accompanying descending GW-signal representing spin-down of a Kerr black hole; and $11\%$ promptly producing a black hole at final coalescence - relatively quiet and dark. {Any possible post-merger signal will have no time delay in prompt formation of a BH-torus system.}

Fig. \ref{fig:009} schematically shows the merger sequence of GW170817-like events which include a diverse MM-afterglow events \citep{Putten2022}.

\subsection{{The GRB211211A-kilonova}} 

EM-GW afterglows may be derived from mergers of an NS with a companion NS, BH or white dwarf (WD). While kilonovae such as AT2017gfo to GW170817 represent the ejection of a few $\%M_\odot c^2$ in neutron-rich matter, considerably less is involved in GRBs derived from ultra-relativistic baryon-poor jets. 

GRB211211A is a case in point given the associated kilonova. 
A WD-NS merger progenitor \citep{Yang2022} can, in principle, be identified by the GW-frequency of its precursor merger chirp. By the relatively large radius of the WD, this frequency is bounded by some four orders of magnitude below that of DNS mergers, i.e., about 0.01\,Hz. Indirect evidence might be obtained from non-detection of a merger chirp, provided the source is within the horizon sensitivity distance of canonical DNS and NS-BH mergers. An associated GRB cannot be ruled out, potentially accompanied by a descending GW-chirp during spin-down of an NS or BH remnant. The degeneracy between the two may be broken by GW-calorimetry, by the ample energy reservoir in angular momentum of a Kerr BH compared to the same in an NS \citep{Putten2022}. 

Perhaps GRB211211A is derived from an NS-BH merger \citep{Rast2022}. In general, however, NS-BH mergers are diverse by expected broad distributions in mass and spin of the BH \citep{Putten1999}. For a BH sufficiently massive and/or slowly spinning ($\gtrsim 28 M_\odot$ for extremal Kerr black hole down to $3.7\,M_\odot$ for a Schwarzschild black hole), it may swallow the NS whole “in one shot” sans tidal break-up, possibly producing a FRB without the usual post-merger drama (Table \ref{T5}). Otherwise, a BH-torus system may form producing a short or long GRB depending on BH-spin \citep{Putten2001,Putten2009}. 

An important open question is a possible diversity in kilonovae powered by (a) relatively high temperature MeV-neutrino winds from an SMNS or HMNS in the immediate aftermath of a DNS merger such as GW170817 or (b) relatively low-temperature MeV-neutrino winds from a BH-torus system (\cite{Putten2003}, \S 6) by its much lower temperature of a few MeV. The simplest scenario for GRB211211A appears to be another DNS merger, whose long duration represents the lifetime of spin of its BH-central engine - like GRB170817A. 

Table \ref{T5} provides a summary of the above providing a prospect of MM-afterglows to CBC involving an NS.

\subsection{{DNS binary masses}} \label{C2}
 
Considering the mass-distribution of NS based on EM-surveys alongside GW170817, GW190425, GRB060614 and GRB211227, this phenomenology suggests that the mass-components and consequently the mass of the system may be decisive in post-merger evolution. 
Fig.\,\ref{fig:7} is illustrative. The total mass of GW170817 was $2.7M_{\astrosun}$ placing it in HMNS category after $10\%$ reduction due to gravitational radiation. This HMNS had a lifetime of about $0.92$\,s \cite[e.g.][]{Murg2021,Putten2019a}. On the other hand, the total mass of GW190425 was $3.4 M_{\astrosun}$ pointing to immediate collapse to BH remnant, i.e., the lifetime of the HMNS formed in the immediate aftermath of the merger was (essentially) zero. 
We conjecture that as the total mass gets larger, the lifetime of the immediate merger product is shorter and, consequently, the probability of a powerful kilonova is 
commensurably smaller (Fig.\,\ref{fig:09}), though a kilonova may still emerge from a black hole-torus system. 
 
\subsection{{Classification of progenitor masses}}
 
Our conjecture suggests that not all DNS mergers successfully produce an SGRB and/or kilonova. To clarify, we categorize the NS mass-distribution to six different regions: $A^{\pm}_{0}$ and $B^{\pm}_{0}$ considering that the mass distribution of NS in DNS systems is within the first peak of the pink part of Fig. \ref{fig:7}. $A$ and $B$ refer to the first and, respectively, second peak. Superscripts $\pm$ represent heaver and lighter masses relative to the central value (``zero") in each peak. Accordingly,
\begin{eqnarray}
\begin{array}{lll}
{\rm GW170817}: & & 2A_0\rightarrow B^+ \\ \\
{\rm GW190425}: & & A_0+B_0 \rightarrow B^+.
\end{array}
\end{eqnarray}
By mass, $B^+$ in the second is at the tail of NS mass distribution, almost at the edge of being a BH. While $2A^-$ results in an SMNS that lives forever and can be the progenitor of NS-driven kilonova, $2B^+$ is expected to produce a BH instantly and possibly a faint 
{torus-driven} kilonova.

Any other combination ends up somewhere in between with opportunities for a full story of MM-afterglows to GW170817-like events or partially so like GRB211227. The latter may include a 
torus-driven kilonova, relatively less energetic than an HMNS-driven kilonova and hence may not be detectable (similar to GW190425). Mid-range events may feature a kilonova similar to AT2017gfo associated with GW170817 but with noticeably weaker/more powerful high-energy EM radiation depending on the total mass of the system.

A similar argument can be applied to CC-SNe event suggesting that the event rate of CC-SNe featuring descending GW-chirps powered by Kerr black holes may be more numerous than their relatively small branching ratio to normal long GRBs.
 
Meanwhile a general classification based on Fig.\,\ref{fig:7}, including the GW surveys, shows that, considering the lower mass threshold of SMNSs to be $1.7M_{\astrosun}$, $46.8\%$ of DNS mergers end up with an SMNS, while $42\%$ produce an HMNS with a finite lifetime giving the opportunity for part of this class to have observable EM counterparts. The remaining $11.2\%$ of DNS mergers have a BH as a prompt remnant, meaning that they would have a silent merger without any (observable) afterglow.


\section{Discussion and Conclusions}\label{Sec:Co}

{The planned LVK O4-5 observations offer unprecedented opportunities to probe new astrophysical processes in extreme transient events in the Local Universe. Detecting new GW-signals requires confirmation cross-correlating output across two (or more) GW-detectors with comparable sensitivities.
Currently, the LIGO H1 and L1 detectors show nominally the same sensitivity. Sensitivity of Virgo is slightly less, while sensitivity of KAGRA in track to become comparable in the near future. GW170817-GRB170817A demonstrates the power of joint detection in EM-GW, opening a radically new window of MM-observations.
We discussed the opportunity of joint GW-EM observations to further our understanding of MM-afterglows to mergers involving a neutron star and CC-SNe. Those leading to the formation of a rapidly spinning Kerr black hole may produce long-duration descending chirps in gravitational radiation.}
 
{GW170817-GRB170817A stands out as a template for what may be expected in MM-afterglows to mergers and, by mass-scaling (\ref{EQN_fGW1}-\ref{EQN_fGW2}), for an outlook on luminous GW-emission from a small fraction of energetic CC-SNe in the Local Universe.
GW170817 was triggered during O2 with the second highest detector yield factor at a relatively close distance of $40 \mbox{Mpc}$ with mid-range mass components (Fig. \ref{fig:7}). These three factors - detector yield factor, close proximity and mid-range mass components - have significant impact on the probability and localization of detecting such event.} 

{A second DNS candidate GW190425 revealed no confirmed EM counterparts. This event, however, was relatively far out (at a distance of about $150\mbox{Mpc}$) and the localization area was significantly wider than the field of view (FOV) of EM observatory.
{Its total mass is considerably larger than that of GW170817}, suggesting a prompt gravitational collapse of the initial HMNS to a black hole-disk or torus system. Non-detection of a kilonova similar to AT2017gfo is therefore not unexpected, also due to relatively inexact localization, 
nor the non-detection of a GRB due to beaming.}

{Non-detection of kilonovae can be challenging by distance. For instance, we recall GRB060614 at $z = 0.125$, attributed to a DNS merger with an associated kilonova, and GRB211227, observed at $z = 0.228$, possibly likewise a DNS merger based on EM-observations \citep{Lue2022} (This event was detected after the completion of O3, however). Although GRB211227 and GRB060614 are similar, the first showed no detectable kilonova. This seems natural, however, as the kilonova associated with GRB060614 was barely observed (one data point) while the redshift of GRB211227 is essentially twice that of GRB060614 which may explain any associated kilonova to be practically undetectable.} 
  
We summarize the results of this study as follows:
\begin{itemize}
\item Discovery power of GW-observations depends critically on true observation time, especially in searches for unknown GW-signals from un-modeled sources, requiring confirmation across two (or more) detectors operating at nominal performance. We quantify this over a choice of time-frequency domain $W\times B$ in terms of detector yield factors of joint H1L1-observations. These yields are also essential to converting observed to true astrophysical event rates, which is crucial when seeking consistency with existing event rates from EM surveys;
\item  Discovery power for unknown GW-signals depends critically on using time-symmetric methods of analysis with equal sensitivity to ascending and descending chirps, anticipated from transient emission in mergers and spin-down of compact objects and/or in the central engine of GRBs and CC-SNe;
\item  DNS mergers may be intrinsically diverse in their MM-afterglow emission. Of particular interest is dependency of mass components (\S\ref{C2}) and, consequently, kilonova characteristics;
\item {Kerr black holes may be universal in powering the most extreme GW-transients such as GRBs and energetic CC-SNe. For O4, the horizon distance to a descending GW-chirp from the latter may be in excess of 100Mpc, based on a relatively low frequency (\ref{EQN_fGW2}) commensurate with $B$ and a relatively large energy output ${\cal E}_{GW}$, implied by a black hole mass appreciably larger than that powering GW170817A.}
\end{itemize}
  
\subsection{Detector yield factors and event rates} \label{C1}

 The true duty cycle of LIGO has an important impact on the MMA related processes and results. The concept of true duty cycle translates to a detector yield factor when both detectors are collecting high-quality data with small standard deviation.
 
 The importance of the detector yield factor is more clear if we consider the relatively high yield factors of EM-observatories \citep[e.g.][]{nasac} and the fact that it is deterministic in estimating the astronomical event rates, i.e.: accurate yield factors are required for consistent event rate in EM-GW surveys. Event rates are crucial in observational astronomy because they are to set the waiting time to capture an event.
 
 {Since LVK is not operating continuously}, precise measurement about their detector yield factors is crucial. A revisit of LIGO runs shows results far different from mere coincidence duty cycles. The coincidence duty cycle of LIGO ranged from $25\%$ to $68\%$ during S5 to O3b, increasing in transition during aLIGO (Table \ref{T1}). Detector yield factors of coincident H1L1-observations (important for MMA and in search of long duration unknown transients) is roughly $1\%$ up to $25\%$ (Table \ref{T2}, Fig. \ref{fig:2}).
 
 Fig. \ref{fig:3} is illustrative for the H1L1-performance during the runs at their most sensitive frequencies, showing the exceptional status of O2. We herein define high quality data as the segments with $\sigma/\mu<1$ and steady phase at the most sensitive frequency of LIGO ($100-250\mbox{Hz}$), in both H1 and L1 and overlapping within $\sigma>0$. {Operating both detectors in high quality state is essential when searching for unknown 
 {GW-signals}, requiring
 confirmation across two independent detectors.}
 
 From a MM point of view, the number of GW-events associated with EM-surveys (Table \ref{Te}) 
 { follows from row 3 of Table \ref{T3}}. 
 These GW-EM survey numbers as they are today provide only a rough 
 estimates of waiting time for a next event. 
 For instance, based on the event rates of SGRB (Table \ref{Te}), GW170817 associated with an SGRB had a probability of about $20\%$ to be detected during O2 (Table \ref{T3}). 
 This has a corresponding estimated event rate of one in 5yr, where time refers to high data quality observing time, at a sensitivity similar to O2 and O3.
 GW190425 with no EM counterpart makes this result more pertinent based on DNS events seen in GWs.
 
 In the previous section, 
 {we estimate potential LVK discovery power for unknown GW-signals by extrapolating performance of current LIGO runs}. According to \eqref{improvement1}, improvement in duty-cycle and data-analysis combined may improve our chances of detecting a transient event of unknown origin by a factor up to 655,000, compared to O3, courtesy of cubic scaling with (realized) sensitivity. Even a fraction thereof, e.g., at realistic yield factors below $100\%$, a detector sensitivity improvement of a factor of $2.5$ by O5 and by employing time-symmetric data-analysis methods (equal sensitivity to ascending and descending chirps), points to improvements of about ${\cal O}\left(10^5\right)$. For the latter, Fig. \ref{fig:TR} is illustrative, which depicts identical results for ascending and descending chirps in a spectrogram of GW170817 by the time-sliced FFT-based MatLab routine {\sc spectrogram}.
 
{To estimate SGRB} event rates using GW-observations, 
{ we can use GW170817-GRB170817A. By \eqref{ov}, during O2 the observational satisfies}
 \onecolumngrid
 \begin{center}
 \begin{equation}
   \label{O2ov}  
   \left < VT \right>=\left <(100\mbox{Mpc})^3 (269\ \mbox{days}) \times 0.45\right >\simeq (100\mbox{Mpc})^3(4\ \mbox{months}).
 \end{equation}
 \end{center}
 \twocolumngrid
 {Given the unique event GRB170817A during O2,} \eqref{O2ov} implies an SGRB event rate of $\sim 3\ \mbox{yr}^{-1} (100\mbox{Mpc})^{-3}$. 
 Taking into account $y\simeq0.19$, the SGRB event rate increases about a factor of two to $7\ \mbox{yr}^{-1} (100\mbox{Mpc})^{-3}$. 
 However, none of these are consistent with $1\ \mbox{yr}^{-1} (\mbox{100Mpc})^{-3}$ from EM-surveys. Instead, it suggests that {\em some DNS systems may merge with no GRB and/or hard to detect sub-luminous EM counterparts.}
 
 Further including O3, 601 operating days, 
 {one GRB170817A and GW190824 having an EM counterpart suggests} an even more modest estimation of the true SGRB event rate. Considering the total coincidence duty cycle of O2 and O3 to be their geometric mean $52\%$, 
 we obtain $1\ \mbox{yr}^{-1} (100\ \mbox{Mpc})^{-3}$,
 apparently consistent with EM-surveys. Given the detector yield factor of about $20\%$, the true event rate is $2-3\ \mbox{yr}^{-1} (100\mbox{Mpc})^{-3}$ - now perhaps above EM event rates. The detector yield factor lowers our expectation for another GW170817-like event in O4, 
 {provided the O4 detector yield factor is similar to O3}. Conversely, if O4 sensitivity is about $160 \mbox{Mpc}$, another GW170817-like event may be detected, provided the detector yield factor is better than $24\%$.
 
 Nevertheless, it appears GW170817 and GW190425 are to be seen jointly for our current survey of DNS events 
 with GW190425 representative for having no EM-counterparts. This supports the idea that not all DNS mergers produce EM afterglows.

\subsection{Relativistic time delay in GRB170817A}

{For DNS mergers, the life-time of HMNS remnants, {dependent on total mass of the system}, is expected to determine diversity of MM-emission that may include a descending chirp in the case of delayed gravitational collapse to a Kerr black hole. 
This delay time, if finite, sets the start time of the central engine of a putative GRB in (\ref{time}). 
By relativistic beaming, an additional delay time $\Delta t$ appears to the observer,} 
 \begin{equation}
     \label{GRB}
     \Delta t = t_{GRB}-t_{s,GW}\approx (0.5\ {\rm s})\ r_{10}\left(\dfrac{\Gamma}{70}\right)^{-2},
 \end{equation}
 {where $r=r_{10} 10\mbox{AU}$, $t_{GRB}$ is the starting time of GRB (associated with GW event), 
 {produced by an ultra-relativistic baryon-poor jet} with Lorentz factor $\Gamma$.} 
 {According to (\ref{GRB}),} more energetic GRBs are detectable in a shorter time interval after their production in the central engine.
 
 {In (\ref{GRB}), for GRB170817A observed is $t_{GRB}-t_m=1.7\,$s.} 
 According to \eqref{time} the life-time of the initial HMNS is $\sim 0.92\,{\rm s}$ marking the birth time of a Kerr black hole central engine. This defines a relativistic time-delay of 
 $\Delta t \sim 0.8\,{\rm s}$ before detecting GRB170817A. 
 Thus, $\Delta t$ is consistent with the recent HST estimate $\Gamma \simeq 70$ \citep{mooley2022}, providing additional consistency between GW and EM-timing highlighted in Table 2 of \cite{Putten2022}. 
 {With $\Gamma$ and $t_{s,GW}$ at hand,} for the first time, we can determine
 \begin{equation}
     \label{r}
     r_{10}=\left(\dfrac{\Delta t}{0.5\,s}\right)\left(\dfrac{\Gamma}{70}\right)^2\approx 1.6,
 \end{equation}
providing a new observational constraint on external shock models for the prompt emission of GRB170817A (cf. \cite{Putten2012}). \eqref{r} is discussed in more detail including off-axis seeing of GRB170817A in \cite{Putten2023a}.

\begin{table*}[]
\begin{tabular}{|c|c|l|ccc|ccc|}
\hline
                          & \multirow{2}{*}{\textit{Merger type}} & \multirow{2}{*}{\textit{Companion}} & \multicolumn{3}{c|}{EM}                                                                                 & \multicolumn{3}{c|}{GW}                                                                                                                                                 \\ \cline{4-9} 
                          &                                       &                                     & \multicolumn{1}{c|}{\textit{FRB}}        & \multicolumn{1}{c|}{\textit{GRB}}       & \textit{Kilonova}  & \multicolumn{1}{l|}{\textit{QNM}}        & \multicolumn{1}{c|}{\textit{descending}} & \multicolumn{1}{c|}{$\mathcal{E}_{GW}$}                          \\ \cline{2-9} 
                          & \multirow{2}{*}{Breakup}              & NS                                  & \multicolumn{1}{c|}{\multirow{2}{*}{-1}} & \multicolumn{1}{c|}{\multirow{2}{*}{1}} & \multirow{2}{*}{1} & \multicolumn{1}{c|}{\multirow{2}{*}{-1}} & \multicolumn{1}{c|}{\multirow{2}{*}{1}}  & \multicolumn{1}{c|}{NS limited}             \\
\multicolumn{1}{|l|}{NS+} &                                       & BH                                  & \multicolumn{1}{c|}{}                    & \multicolumn{1}{c|}{}                   &                    & \multicolumn{1}{c|}{}                    & \multicolumn{1}{c|}{}                    & \multicolumn{1}{c|}{KBH limited}                                                \\ \cline{2-9} 
                          & One-shot                              & BH                                  & \multicolumn{1}{c|}{1}                   & \multicolumn{1}{c|}{-1}                 & -1                 & \multicolumn{1}{c|}{1}                   & \multicolumn{1}{c|}{-1}                  & \multicolumn{1}{c|}{1}                                       \\ \cline{2-9} 
                          & \multirow{2}{*}{Sweet}                & \multirow{2}{*}{WD}                 & \multicolumn{1}{c|}{\multirow{2}{*}{-1}} & \multicolumn{1}{c|}{\multirow{2}{*}{0}} & \multirow{2}{*}{0} & \multicolumn{1}{c|}{\multirow{2}{*}{-1}} & \multicolumn{1}{c|}{\multirow{2}{*}{0}}  & \multicolumn{1}{c|}{NS limited}      \\
\multicolumn{1}{|l|}{}    &                                       &                                     & \multicolumn{1}{c|}{}                    & \multicolumn{1}{c|}{}                   &                    & \multicolumn{1}{c|}{}                    & \multicolumn{1}{c|}{}                    & \multicolumn{1}{c|}{KBH limited}                                               \\ \hline
\end{tabular}
\caption{{High-energy MM-afterglows to CBC involving an NS. {\it Breakup mergers} are those forming a BH-torus system. When initially rapidly spinning, this may produce a descending chirp during BH spin-down powering an associated GRB of similar duration and, possibly, a kilonova. {\it One-shot mergers} represent CBC in which the BH swallows the NS whole in one-shot sans breakup, when the BH is sufficiently massive and/or spins slowly. The result may be a FRB without the usual post-merger drama such as GW170817. {\it Sweet mergers} refer to NS-WD mergers. While expected to be more frequent than DNS mergers, their coalescence is hard to detect by GW-frequencies some four orders of magnitude below that of NS-NS or NS-BH mergers. Here, ``-1'', ``0'' and ``1'' indices represent “impossible”, “possible but without observational evidence”, “possible with observational evidence”, respectively. QNM refers to Quasi-Normal Mode ringing of the event horizon of the black hole. }}
\label{T5}
\end{table*}

\subsection{Advances in GW over EM$\nu$ observations}
{GW170817 is a breakthrough MM-event by the associated GRB170817A and kilonova AT2017gfo. 
The key open question is the evolution of the initial HMNS remnant and the ensuing trigger of the birth of the central engine of GRB170817A. This poses inevitably the challenge to break the degeneracy between this NS and the BH that may form by delayed gravitational collapse for which the EM observations are inadequate \citep{Lazz2020,Piro2019,Metz2018,Pool2018}:}

\begin{itemize}
    \item {According to (\ref{GRB}-\ref{r}), the prompt GRB emission of GRB170817 originates at a distance of about 16\,AU away from the source, inhibiting a direct view on its central engine. }

    \item {The total energy radiated in the EM channel is $\mathcal{E}_{EM}\simeq 0.5\%M_{\odot}c^2$ \citep{mooley2022}. This is about five times smaller than the energy $\mathcal{E}_{GW}\simeq 2.5\%M_{\odot}c^2$ radiated in GW170817: {\it EM radiation is vastly subdominant.}}

    \item {While the kilonova AT2017gfo is representative of neutron-rich mass ejecta largely from the HMNS, it perhaps includes contributions from the BH-turos system formed following {a potential delayed gravitational collapse} of the HMNS to a Kerr BH given the estimated mass of $\sim 2.4M_{\odot}$. If so, the inferred life-time of the  HMNS \citep{Gill2019} {reduces to an upper bound.}}

\end{itemize}

{An additional neutrino detection - which is absent for GW170817 - is not likely to change these conclusions. After all, SN1987A revealed a dominant neutrino channel, yet its remnant is still unknown \citep{Alp2018}.}

Note that this is not so for GWs. A revisit of data around GW170817, detects a {descending chirp (Figs. \ref{fig:5} and \ref{AA})} at $5.5 \sigma$ C.L. signaling the delayed gravitational collapse of HMNS to a Kerr BH. {This interpretation derives from the energy $\mathcal{E}_{GW}\simeq 3.5\%M_{\odot}c^2$ making GW emission the dominant radiation channel, over the frequency range \eqref{EQN_fGW1}. By the start-time of \eqref{time}, it represents the trigger leading to GRB170817A \citep{Putten2022,Putten2003}. {As elucidated in Fig. \ref{fig:11}, this signal would not be detectable by earlier LIGO searches \citep{Abbott2017e} in light of the sensitivity threshold of $650\%M_\odot c^2$ exceeding the total mass-energy of the system of about $270\%M_\odot c^2$ \citep{Sun2019}.}}

\subsection{THESEUS: a next-generation GRB mission}

The MM-afterglow to GW170817 - a descending branch in GW-emission, GRB170817 and kilonova AT2017gfo - marks a complex transition of the DNS progenitor involving an HMNS with delayed gravitational collapse to its BH remnant. As indicated in Fig. \ref{fig:09}, GW170817 may be part of a one-parameter family of DNS mergers in which the massive NS formed in the immediate aftermath of the merger shows different lifetimes ranging from infinity to zero, marking the formation of an SMNS and, respectively, prompt collapse to a BH.

Furthermore, GW170817 EM afterglow is a subdominant radiation channel with $\mathcal{E}_{EM}=0.5\%M_{\astrosun}c^2 \ll 2.5\%M_{\astrosun}c^2 $ (Fig. \ref{fig:11}, left panel). For SN1987A the dominant radiation channel is MeV-neutrinos with somewhat similar energy output $\mathcal{E}_{\nu}\simeq 4\%M_{\astrosun}c^2$ \citep{Hira1987}.  Thus, GW170817 and SN1987A demonstrate dominant emission channels beyond EM-radiation. For total calorimetry, evidently, a comprehensive EM-GW survey seems pertinent.

A statistically significant sample of MM-observations may resolve the anticipated diversity in DNS merger afterglow phenomenology, including NS-BH events whose post-merger evolution lacks an (intermediate) massive NS. 
Realizing an EM-GW survey of DNS and NS-BH mergers promises to unambiguously identify post-merger evolution, derived from total calorimetry across all radiation channels and detailed event timing in both EM-GW radiation channels. 
In fact, consistency in EM-GW timing will be crucial to confirm identification of the source and central engine of an associated kilonova and (short) GRB (the latter may further be with or without extended emission, known to satisfy the Amati-relation of LGRBs \cite[e.g.][]{Norr2006,Amati2002,Amati2006,Putten2014}). 
Notably this applies to determining the total contribution of merger-induced kilonovae to the cosmic abundances of the heavy elements produced in the r-process. 

Significant prospects to realize such surveys are expected from next-generation GRB space mission concepts, currently being developed and proposed by large international collaborations, notably {\sc THESEUS} ({\em Transient High-Energy Sky and Early Universe Surveyor}) \citep{Amati2018,Amati2021} and the {\sc Gamov Explorer} \citep{White2021}. These projects aim at overcoming some of the current limitations in GRB detection, localization, characterization and multi-wavelength follow-up by extending GRB monitoring down to soft X-rays and performing autonomous near infrared (NIR) follow-up, arcsec localization and redshift measurement. 

In particular, an unprecedented GRB and transient surveyor like {\sc THESEUS} would be capable of providing a substantial advancement in the aforementioned observational dimensions across all classes of GRBs \citep{Stra2018,Ciol2021} with particular emphasis on events at high-$z$, on- and off-axis (such as GRB170817A), short GRBs with associated GW-emission, under-luminous GRBs associated with CC-SNe, which may nevertheless feature detectable GW-emission from their central engines, and the like. 

For short GRBs, the {\sc THESEUS} {\em X-Gamma ray Imaging Spectrometer} (XGIS, 2 keV-20 MeV) and {\em Soft X-ray Imager} (SXI, 0.3-5 keV) may detect and localize $\sim 40\,$ events at $<\,15\,$arcmin (90\%) and $<\,7\,$arcmin (50\%) events over a 3.5\,yr time of observation, accounting for all payload observational constraints. 
At higher energies ($>\,150$\,keV), short GRBs are expected to be detected at a rate of $\sim 28$/yr outside the FOV of XGIS. While at relatively course sky localization ($\sim 500\,$deg$^2$), this nevertheless may combine with GW-observations by temporal coincidences.

{\sc THESEUS} is ideally equipped to provide accurate redshifts from detections at low energy from optical/IR afterglow synchrotron emission in the propagation of the short GRB jet, propagating into the interstellar medium (ISM). 
A fair percentage of prompt short GRB detections ($\sim 3$/yr) are expected to be accompanied by Infrared Telescope (IRT) afterglow detection by {\sc THESEUS} {\em InfraRed Telescope} (IRT covering 0.7-1.8$\mu$m with a 0.7\,m class IR-telescope over an FoV 15$^\prime \times 15^\prime$ with spectroscopic capabilities). 
Accurate arcminute sky localization by {\sc THESEUS}-XSI with, where available, IRT, will allow facilities such as {\sc Athena} \citep{Nand2013}, {\sc E-ELT} \citep{Gilm2007} and the SquareKilometreArray ({\sc SKA}) \citep[e.g.][]{Jans2015} to deeply monitor and further characterize the source, significantly improving chances to identify the host galaxy with redshift. Conservative estimated point to a yield of $>26$, assuming follow-up by other facilities within hours. 

Notable is that mere duration of GRBs cannot be representative for all the properties of these relativistic transients \citep{Amati2021}. A fraction of technically long GRBs may be produced by DNS mergers and be classified as SGRB with Extended Emission (SGRBEE). GRB211211A may just be such an SGRBEE at a distance of 346Mpc with a duration of about 50 seconds, with an associated kilonova similar to the one during
GW170817 \citep{Troja2022,Mei2022,Rast2022,Yang2022}. Such might also be derived from some of the NS-BH or WD-NS mergers (Table \ref{T5}, \S\ref{C2}). This outlook, taking us beyond SGRBs alone, suggests a possible increase in rates of joint GW-GRB detection in joint LVK-THESEUS operation.

For DNS mergers, these redshift determinations supplementing accurate measurement of chirp mass and associated GW-luminosity provide a potentially powerful means for a determination of the Hubble parameter in late-time cosmology \citep[e.g.][]{Abbott2017h,Guid2017}, independent of the Local Distance Ladder. Such may be crucial in confirming the present $H_0$-tension between the latter and what is inferred from Planck $\Lambda$CDM analysis of the CMB \citep{Adam2022}. 
Above-mentioned EM-GW surveys promise detailed parameterization of MM-afterglows to DNS events, which might lead the way to normalize descending branches in GW-emission, possibly improving  the potential of DNS mergers as standard sirens of GW-emission. A decisive determination of the Hubble parameter in late-time cosmology may provide us with a probe of (in)stability of de Sitter space in low-energy quantum cosmology \citep[e.g.][]{Putten2021}. 

\section*{Acknowledgment}
We are thankful to the anonymous reviewer for providing insightful comments on this manuscript. The authors are also grateful to Cristiano Guidorzi for his constructive comments on this manuscript, specifically on event rates, and to Peter Hoeflich for useful information about r-process and heavy elements abundances. This research is supported by NRF of Korea Nos. 2018044640. M. van Putten thanks LIGO-Caltech for hospitality over a sabbatical visit, where some of this work was performed. L. Amati acknowledges support from the Italian Ministry of Research through grant PRIN MIUR 2020 – 2020KB33TP METE. This research has made use of S5-O3ab data obtained from the Gravitational Wave Open Science Center (gw-openscience.org), a service of LIGO Laboratory, the LIGO Scientific Collaboration, the Virgo Collaboration, and KAGRA. LIGO Laboratory and Advanced LIGO are funded by the United States National Science Foundation (NSF) as well as the Science and Technology Facilities Council (STFC) of the United Kingdom, the Max-Planck-Society (MPS), and the State of Niedersachsen/Germany for support of the construction of Advanced LIGO and construction and operation of the GEO600 detector. Additional support for Advanced LIGO was provided by the Australian Research Council. Virgo is funded, through the European Gravitational Observatory (EGO), by the French Centre National de Recherche Scientifique (CNRS), the Italian Istituto Nazionale di Fisica Nucleare (INFN) and the Dutch Nikhef, with contributions by institutions from Belgium, Germany, Greece, Hungary, Ireland, Japan, Monaco, Poland, Portugal, Spain. The construction and operation of KAGRA are funded by Ministry of Education, Culture, Sports, Science and Technology (MEXT), and Japan Society for the Promotion of Science (JSPS), National Research Foundation (NRF) and Ministry of Science and ICT (MSIT) in Korea, Academia Sinica (AS) and the Ministry of Science and Technology (MoST) in Taiwan.

\appendix
\counterwithin{figure}{section}
\counterwithin{table}{section}
\section{Observations by LIGO} \label{Sec:GWOL}

{With the first direct detection of gravitational waves GW150914, LIGO opened a radically new window of observation to the Universe.
The LIGO project has evolved through various generations of engineering, science and observational runs since 1995.} A primary objective has been to detect gravitational waves from astrophysical {transients events sources with rigorous confirmation across the two independent detectors at Hanford (H1) and Livingston (L1)}. LIGO is now jointly operated with Virgo (EU),  KAGRA (Japan), GEO600 (Germany)\citep{Somi2012,Aso2013,Abbott2020,Schu2011,Abbott2008b}, expected to include IndIGO in the near future \citep{Fair2014,Rile2013}. 

LIGO has performed six scientific runs \citep{Abbott2009a,Abbott2008,Abbott2009b,Abbott2009,Aasi2014,Abbott2008a}. 
During S5 (4/11/2005-1/10/2007), {H1 and L1 had duty cycles of $78\%$ and, respectively, $67\%$ with a coincidence duty cycle of $60-68\%$ \citep{Abbott2009,Abad2010}.} 
Duty cycle {refers to} the fraction of total run time with science quality data \citep{Aasi2015}. 
For a fiducial $1\%$ output in gravitational radiation in a CBC relative to the total mass-energy of the system, the horizon distance of S5 for DNS was estimated to be $\sim 30-35 \mbox{Mpc}$, while for binary black holes (BBH) of $M_1=M_2=5 M_{\astrosun}$ it was about $\sim 100 \mbox{Mpc}$ \citep{Collab2010,Abad2010}. 
S6 {(7/7/2009-20/10/2010) - the last run before Advanced LIGO - shows} improvements in sensitivity at a reduced duty cycle (Fig. \ref{fig:2}, Table \ref{T1}). 
Duty cycles of H1 and L1 during S6 are $50\%$ and, respectively, $47\%$, giving a simultaneous observing time of 117 days ($25\%$). The S6 horizon distance is estimated to be $\sim 35-45 \mbox{Mpc}$ for equal mass binary neutron stars \citep{Abad2012}, and $\sim 300 \mbox{Mpc}$ for BBH of $M_1=M_2=20 M_{\astrosun}$ \citep{Aasi2013}. {Nevertheless, no event candidates have been identified during S5 and S6.}

{During O1 of Advanced LIGO (aLIGO)}, GW150914 presented the first-ever direct detection of gravitational waves. The detectors were three to five times more sensitive in the frequency range of $100-300 \mbox{Hz}$ compared to initial LIGO (iLIGO) prior \citep{Abbott2017,Abbott2017c}.
{O1 extends over 130 observational days \cite{Abbott2016} (12/09/2015-19/01/2016)}. 
Detailed data-analysis shows GW150914 to be a {BBH merger with masses $M_1=35^{+5}_{-3}M_{\astrosun}$ and $M_2=30^{+3}_{-4}M_{\astrosun}$ at a luminosity distance of $440^{+160}_{-180} \mbox{Mpc}$ \citep{Abbott2016c}. }
The remnant of this merger is a black hole of mass $M=62^{+4}_{-3}M_{\astrosun}$.
{It indicates a total energy release $\mathcal{E}_{GW}\cong 3.0 \pm 0.5 M_{\astrosun} c^2$ in gravitational waves \citep{Cast2016,Abbott2016d}, i.e., a few percent of the total mass-energy of the system, confirmed by similar events in subsequent observational runs \citep{Aasi2014}.} 
Following a normalization to the unit of luminosity 
\begin{eqnarray}
L_0=\frac{c^5}{G} \simeq 200,000\ M_{\astrosun} c^2/\mbox{s}
\label{EQN_L0}
\end{eqnarray}
of general relativity, where $c$ is the velocity of light and $G$ is Newton’s constant, GW150914 features a peak GW-luminosity of about 
\citep{Putten2019} 
\begin{eqnarray}
\hat{L}_{GW} \simeq 0.1\%L_0
\end{eqnarray}
characteristic for BBH events.

Run O1 added a few more BBH coalescence events to the transient catalog of LIGO. The O1 duty cycle of H1 was $62-70\%$ \citep{Abbott2016a,Buik2020}. The observing time of L1 covers $55\%$ of the total time \citep{Abbott2016a}, giving a coincidence duty cycle of $40-48\%$ \citep{Abbott2016b,Buik2020}. 
{The O1 horizon distance was $\sim 1.3 \mbox{Gpc}$ and $~70-80 \mbox{Mpc}$ for BBH similar to GW150914 and, respectively, DNS mergers \citep{Mart2016}.}
This 
{clearly} evidences aLIGO improvement over iLIGO. The spectral density of LIGO detectors during O1 was $\sim 10^{-23} \, \sqrt{1/\mbox{Hz}}$ \citep{Abbott2016}. An improvement by a factor of 3 was expected for the next observational runs \citep{Abbott2017}.

{By improved sensitivity and spectral density, O2 showed a horizon distance of $70-100 \mbox{Mpc}$ for DNS and $\sim 8 \times 10^{-24}\, \sqrt{1/\mbox{Hz}}$ \citep{Davis2021,Mart2016}.} 
O2 extended over 269 days (30/11/2016-25/08/2017) \citep{Abbott2017b} with an H1 duty cycle of $60-70\%$ \citep{Buik2020,Davis2021,Abbott2019,Abbott2021} and an L1 duty cycle of $60-65\%$ \citep{Davis2021,Buik2020,Abbott2019,Abbott2021}, leaving a joint H1L1-duty cycle of $45-46\%$ \citep{Abbott2019,Buik2020}. 

{During O2, LIGO successfully detected GW170817: the first DNS coalescence event with a short gamma ray burst GRB170817A, detected by INTEGRAL and \textit{Fermi}-GBM 1.7 seconds after the coalescence. LIGO and \textit{Fermi}-GBM detections at the same location} {in the sky and time provides evidence for a secure association of GRB170817A with GW170817 at} a significance of $4.8 \sigma$. This marks GW170817-GRB170817A as a genuine multi-messenger event {further by the associated kilonova AT2017gfo} \citep{Abbott2017d,Abbott2017b,Abbott2017a,Gold2017}. 
GW170817 was detected at the luminosity distance of $40^{+8}_{-14} \mbox{Mpc}$. This is quite {fortuitous} considering the DNS merger rate of $320^{+490}_{-240} \mbox{Gpc}^{-3}\mbox{yr}^{-1}$ obtained from EM surveys \citep{Abbott2017b,Abbott2017d}. {It should be mentioned, however, that estimating event rates in the Local Universe by extrapolation of rates on cosmological scales tends to be notoriously uncertain.}

Although there has been some discussion about the binary GW170817 \citep{Rued2018,Coug2019,Tsai2021}, a DNS is the main candidate \citep{Putten2019a,Putten2022,Coug2019,Abbott2019b}. The alternative of an NS-BH is effectively ruled out by the delayed collapse of an HMNS in the initial aftermath of this merger (\cite{Putten2022}; Fig. \ref{fig:009}). GW170817 is a most exceptional event also for {a first-ever demonstration of measurement of the Hubble constant $H_0$ by a DNS merger \citep{Abbott2017e,Abbott2017f,Fish2019}, independent of Planck-$\Lambda$CDM and the Local Distance Ladder. {O2 increased the total number of detected CBCs by LVK to 11 events \citep{Abbott2019}. Two additional events were reported later by \cite{Zack2021}.}}

The most recent LIGO run O3 - O3a (01/04/2019-01/10/2019, 184 days) and O3b (01/11/2019-27/03/2020, 148 days) \citep{Buik2020} - extended over 11 months. O3 represents some notable improvements in duty cycles and sensitivity \citep{Davis2021,Abbott2019a}. During O3a, the duty cycles of H1 and L1 are $71\%$ \citep{Buik2020,Abbott2021} and, respectively, $76\%$ \citep{Buik2020,Abbott2021}. For O3b, H1 and L1 have duty cycles of $78\%-79\%$ \citep{Buik2020,Collab2021}. Detector sensitivity improved by a factor of two over O2, i.e. to $\sim 4 \times 10^{-24}$ \citep{Davis2021}. Accordingly, the BBH horizon distance for $M_1=M_2=30M_{\astrosun}$ increased to $1.15 \mbox{Gpc}$ for H1 and $1.425 \mbox{Gpc}$ for L1 \citep{Davis2021}. 
For DNS, the O3ab horizon distances are $108-115 \mbox{Mpc}$ for H1 and $133-135 \mbox{Mpc}$ for L1 \citep{Buik2020,Abbott2021,Collab2021}. 
Table \ref{T1} lists some statistics of duty cycles, coincidence duty cycles ($\mathcal{U}$), sensitivity and horizon distances. These continuous improvements {have} produced an extensive LIGO-Virgo catalog of CBC events \citep{Collab2021,Collab2021a,Abbott2019,Abbott2021}. {This catalog was updated later to include 7 additional BBH \citep{Nitz2021}.}
\begin{center}
	\begin{table*}[]
		\footnotesize
		\begin{tabular}{|l||l|l|l|l|l|l|}
		\hline
		\hspace{0.1cm}	& Total  & Individual  & Coincidence  & Sensitivity &  Horizon  &  Horizon  \\ 
		Run	&  number of  & duty cycles & duty cycles ($\mathcal{U}$)& (spectral density) &  distance BBH &  distance DNS \\ 
		\hspace{0.1cm}	& days & [$\%$] & [$\%$] & [$\sqrt{1/Hz}$] &  [Mpc] &  [Mpc] \\ \hline \hline
		&    &  &  &  &    &     \\
		S5	& $697^{(1)}$  & H1: 78  & $60^{(1)}$ & $\sim 3 \times 10^{{-22}^{(1)}}$ &  3 ($10 M_{\astrosun}$)$^{(1)}$ &  30-35 $^{(2,3,4)}$ \\ 
		\hspace{0.1cm}	&  \hspace{0.1cm}  & L1: 67 $^{(1)}$ & 68 $^{(4)}$ & \hspace{0.1cm} &  90 $^{(4)}$ &  \hspace{0.1cm} \\
		\hspace{0.1cm}	&  \hspace{0.1cm}  & \hspace{0.1cm} & \hspace{0.1cm} & \hspace{0.1cm} &  100 ($5 M_{\astrosun}$) $^{(3)}$ &  \hspace{0.1cm} \\
		\hspace{0.1cm}	&  \hspace{0.1cm}  & \hspace{0.1cm} & \hspace{0.1cm} & \hspace{0.1cm} &  120 ($50 M_{\astrosun}$) $^{(1)}$ &  \hspace{0.1cm} \\ \hline
		&    &  &  &  &    &     \\
		S6	&471 $^{(5)}$  & H1: 50  & 25 $^{(6)}$ & $\sim 2 \times 10^{{-23}^{(6)}}$ &  300 ($20 M_{\astrosun}$) $^{(7)}$ &  40 $^{(8)}$ \\ 
			&    & L1: 47 $^{(9)}$ &   &   &  90 ($5 M_{\astrosun}$) $^{(5)}$ &  35-45 $^{(10)}$ \\ \hline
			&    &  &  &  &    &     \\
		O1	&130 $^{(11)}$  & H1: 62-70 $^{(12,13,14)}$ & 40-48 $^{(12,14,15)}$ & $\sim 8\times 10^{-24}$ $^{(11,16)}$ &  1300 ($30 M_{\astrosun}$) $^{(16)}$ &  70-80 $^{(16)}$ \\
			&     & L1: 55-57 $^{(13,14)}$ &   &   &    &     \\ \hline
			&    &  &  &  &    &     \\
		O2	&269 $^{(17)}$  & H1: 60-70  & 45 $^{(14,18)}$ & $\sim 7 \times 10^{-24}$ $^{(19)}$ &  Farthest event &  70-100 $^{(18)}$ \\
			&    & L1: 60-65  &  &   &: 2840 $^{(18)}$   &   \\ 
			& &$^{(11,12,14,19)}$ & & & & \\ \hline
			&    &  &  &  &    &     \\
		O3a	&184 $^{(12)}$  & H1: 71  & 62 $^{(12)}$ & $\sim 5 \times 10^{-24}$ $^{(19)}$ &  H1: 1150 ($30 M_{\astrosun}$) &  H1: 108,111  \\
			&    & L1: 75-76 $^{(12,20)}$ &  &  &  L1: 1425 ($30 M_{\astrosun}$)  &  L1: 134,135  \\ 
			& & & & & $^{(19)}$&$^{(12,20)}$ \\ \hline
			&    &  &  &  &    &     \\
		O3b	&148 $^{(12)}$  & H1: 78-79  & 62 $^{(12)}$ & $\sim 5 \times 10^{-24}$ $^{(19)}$ &  H1: 1150 ($30 M_{\astrosun}$) &  H1: 115  \\
			&    & L1: 78-79 $^{(12,21)}$ &  &  & L1: 1425 ($30 M_{\astrosun}$)  &  L1: 133 $^{(21)}$  \\ 
			&    &  &  &  & $^{(19)}$    &     \\ \hline
			
		\end{tabular}
		\caption{\small
		\textit{Summary of some properties of LIGO runs}. Listed are statistical results of S5, S6, O1, O2, O3ab. The H1-L1 detectors have undergone major improvements during 16 years of operation, most importantly migrating to aLIGO from iLIGO in sensitivity. The second column shows the total number of days covered by each run. For instance S5 extended over 697 days (04.11.2005-01.10.2007). Individual duty cycles of H1 and L1 are presented in the third column, indicating the fraction of time of operation and in taking science data \citep{Abbott2009}.  {Times when operations overlap} defines \textit{coincidence duty cycles}, included in the fourth column. A slight improvement in duty cycles {is noticeable} during aLIGO, starting from O1. The fifth column presents LIGO sensitivity during each run based on the spectral density of the strain ($\sqrt{S_n(f)}$) and the improvement therein. {Sensitivity improvements can be expressed by horizon distance out to which can explore a standard event (BBH or DNS merger) or, equivalently, by detectors sensitivity defined by strain-noise amplitude (columns 6-7). Values are calculated for standard events mentioned in parentheses.}}
	$^{(1)}$\cite{Abbott2009};$^{(2)}$\cite{Collab2010};$^{(3)}$\cite{Abbott2009a};$^{(4)}$\cite{Abad2010};$^{(5)}$\cite{Abad2012};$^{(6)}$\cite{Aasi2015};$^{(7)}$\cite{Aasi2013};$^{(8)}$\cite{Abad2012};$^{(9)}$\cite{Aasi2014};$^{(10)}$\cite{Collab2012};$^{(11)}$\cite{Abbott2016};$^{12}$\cite{Buik2020};$^{13}$\cite{Abbott2016a};$^{(14)}$\cite{Abbott2020},$^{(15)}$\cite{Abbott2016b};$^{(16)}$\cite{Mart2016};$^{(17)}$\cite{Abbott2017b};$^{(18)}$\cite{Abbott2019};$^{(19)}$\cite{Davis2021};$^{(20)}$\cite{Abbott2021};$^{(21)}$\cite{Collab2021}
		\label{T1}
	\end{table*}
\end{center}

\section{Time-symmetric analysis method} \label{Sec:TSAM}
{An important component in studying transients data is employing a method of analysis with equal sensitivity to ascending and descending signals: a time-symmetric method. This is essential in un-modeled searches of GW signals from unknown sources. A prototype of time-symmetric method is time-sliced FFT showing identical results for the strength of a merger signal at each frequency in comparison with its time reversal, illustrated in Fig. \ref{fig:TR}. Optimal performance of time-sliced FFT depends on the density of the filter in each frequency. But FFT is most sensitive to narrow band frequencies. Specifically the time-frequency uncertainty equation $\Delta t \Delta f \simeq 1$ puts restrictions on the capability of this method in detecting signals with frequencies varying in time over an interval of phase-coherence $\Delta t$. Butterfly matched filtering is a time-symmetric method of analysis, going one step further, that covers the time-varying frequencies with a filter dense in both $f$ and $\dot{f}$ \citep{Putten2014a,Putten2023a}. }

\begin{figure*}[!htb]
\center{\includegraphics[scale=0.43]{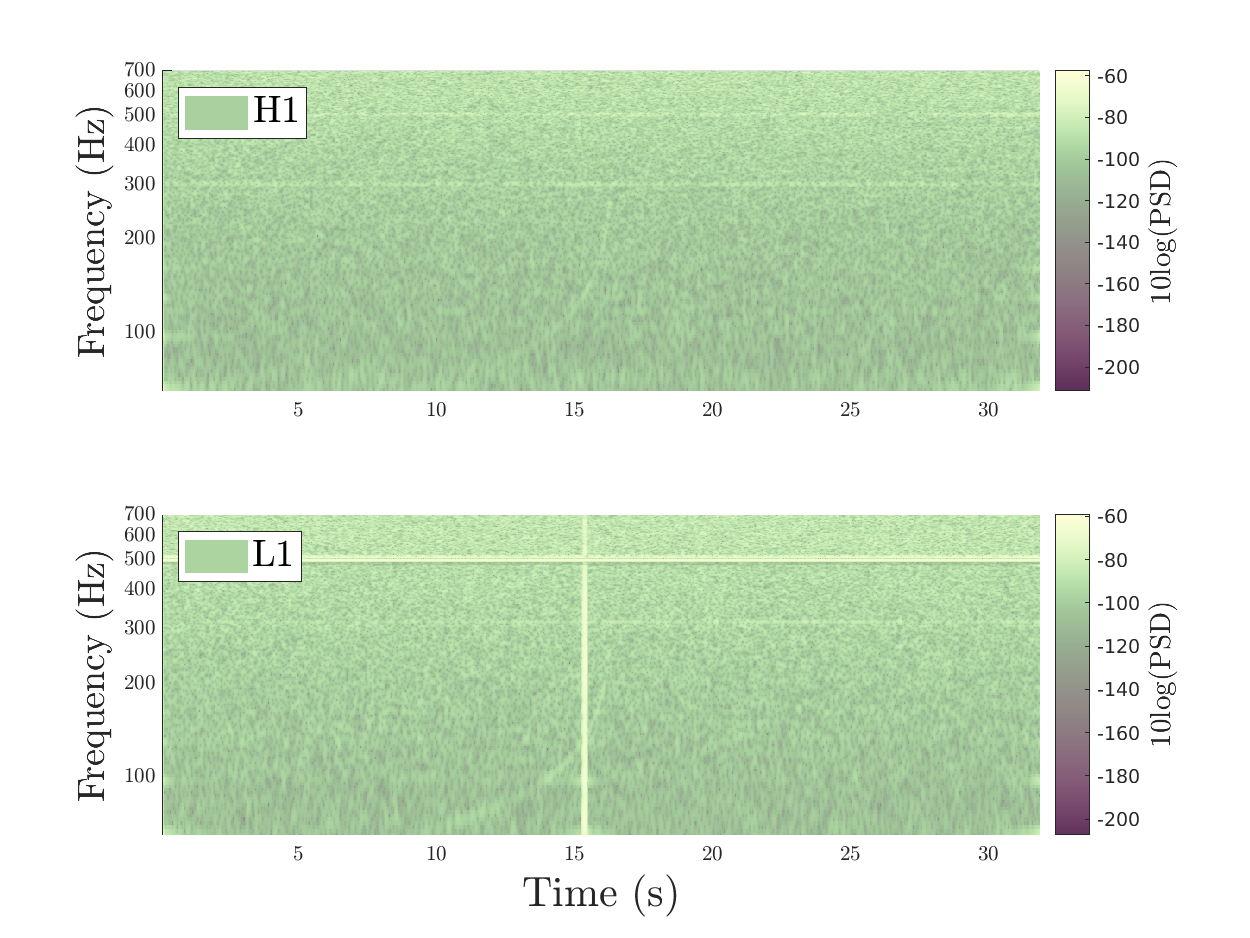}\includegraphics[scale=0.43]{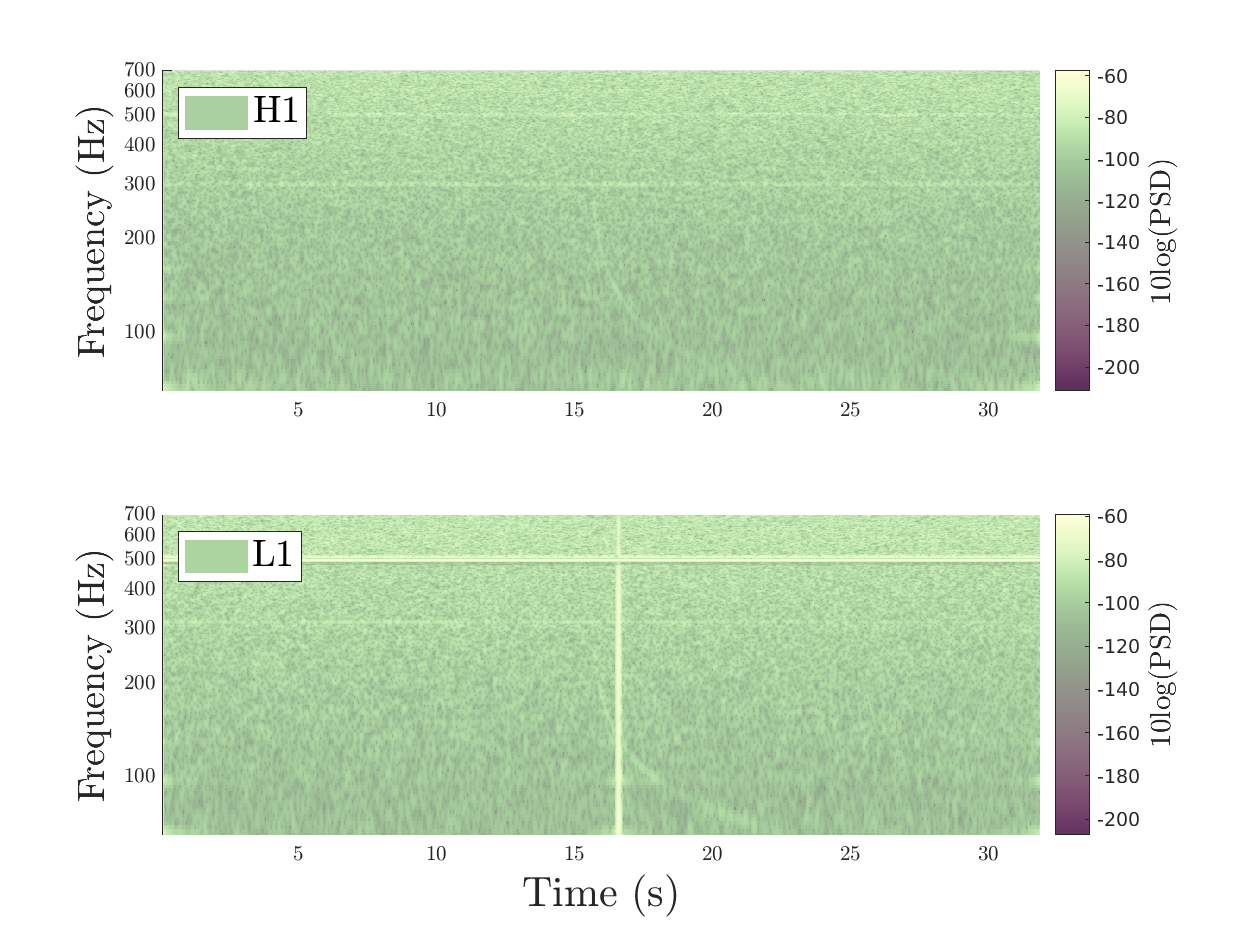}}			
\caption{{\small (Left panel.) The spectrogram of 32\,s of data around GW170817 
{produced by GWXplore, a user-friendly laptop App employing conventional time-sliced FFT}. Here, the FFT window size is set to be $0.25\,{\rm s}$, whitening bandwidth $B=4\mbox{Hz}$ and low-pass and high-pass frequencies are $10\,\mbox{Hz}$ and $1700\,\mbox{Hz}$, respectively. 
(Right panel.) FFT-based analysis of time-reversed data produces the same, demonstrating time-symmetry in the analysis relevant to targeting ascending and descending chirps from CBC including the potential 
{descending emission from the spin-down of merger remnants and central engines of CC-SNe.}
We have made partial use of the color maps in \cite{Thyng_2016}.}}
\label{fig:TR}
\end{figure*} 

\bibliographystyle{aasjournal}
\bibliography{mybibfile}



\end{document}